\documentclass[fleqn,usenatbib]{mnras}

\usepackage{newtxtext,newtxmath}

\usepackage[T1]{fontenc}
\usepackage{ae,aecompl}

\usepackage{graphicx}	
\usepackage{amsmath}	
\usepackage{amssymb}	
\usepackage{threeparttable}

\newcommand{\teq}{$T_\mathrm{eq}$}
\newcommand{\tirr}{$T_\mathrm{irr}$}
\newcommand{\myemail}{giovanni.bruno@inaf.it}
\def\gtsima{$\; \buildrel > \over \sim \;$}
\def\ltsima{$\; \buildrel < \over \sim \;$}
\def\gtrsim{\lower.5ex\hbox{\gtsima}}
\def\lesssim{\lower.5ex\hbox{\ltsima}}

\title{WASP-52$\lowercase{b}$. The effect of starspot correction on atmospheric retrievals}

\author[G. Bruno et al.]{
Giovanni Bruno,$^{1,3}$\thanks{E-mail: \myemail}
Nikole K. Lewis,$^{2,3}$
Munazza K. Alam,$^{4}$
Mercedes L\'opez-Morales,$^{4}$
\newauthor
Joanna K. Barstow,$^{5}$
Hannah R. Wakeford,$^{3}$
David Sing,$^{6}$
Gregory W. Henry,$^{7}$
\newauthor
Gilda E. Ballester,$^{8}$
Vincent Bourrier,$^{9}$
Lars A. Buchhave,$^{10}$
Ofer Cohen,$^{11}$
\newauthor
Thomas Mikal-Evans,$^{12}$
Antonio Garc\'ia Mu\~noz,$^{13}$
Panayotis Lavvas,$^{14}$
Jorge Sanz-Forcada$^{15}$
\\
$^{1}$INAF - Catania Astrophysical Observatory, Via Santa Sofia, 78, 95123, Catania, Italy\\
$^{2}$Department of Astronomy and Carl Sagan Institute, Cornell University, 122 Sciences Drive, Ithaca, NY 14853, USA\\
$^{3}$Space Telescope Science Institute, 3700 San Martin Drive, Baltimore, MD 21218, USA\\
$^{4}$Center for Astrophysics | Harvard \& Smithsonian, 60 Garden Street, Cambridge, MA 02138, USA\\
$^{5}$University College London, Department of Physics and Astronomy, United Kingdom\\
$^{6}$Physics and Astronomy, Jonhs Hopkins University, Baltimore, MD, United States\\
$^{7}$ Center of Excellence in Information Systems, Tennessee State University, Nashville, TN 37209, USA\\
$^{8}$Lunar and Planetary Laboratory, University of Arizona, Tucson, AZ 85721, USA\\
$^{9}$Observatoire de l'Universit\'e de Gen\`eve, Chemin des Maillettes 51, Versoix, CH-1290, Switzerland\\
$^{10}$ DTU Space, National Space Institute, Technical University of Denmark, Elektrovej 328, DK-2800 Kgs. Lyngby, Denmark\\
$^{11}$ Lowell Center for Space \& Technology, 600 Suffolk St., Lowell MA 01854, USA\\
$^{12}$ Kavli Institute for Astrophysics and Space Research, Massachusetts Institute of Technology, 77 Massachusetts Avenue, 37-241, Cambridge, MA 02139, USA \\
$^{13}$ Zentrum f\"ur Astronomie und Astrophysik, Technische Universit\"at Berlin,
Berlin, Germany\\
$^{14}$ GSMA, Reims Champagne-Ardenne, F-51687 Reims, France\\
$^{15}$ Centro de Astrobiolog\'{i}a (CSIC-INTA), ESAC Campus, Camino bajo del Castillo s/n, E-28692 Villanueva de la Ca\~nada, Madrid, Spain}

\date{Accepted XXX. Received YYY; in original form ZZZ}

\pubyear{2019}

\begin{document}
\label{firstpage}
\pagerange{\pageref{firstpage}--\pageref{lastpage}}
\maketitle

\begin{abstract}
We perform atmospheric retrievals on the full optical to infrared ($0.3-5 \, \mu \mathrm{m}$) transmission spectrum of the inflated hot Jupiter WASP-52b by combining \textit{HST}/STIS, WFC3 IR, and \textit{Spitzer}/IRAC observations. As WASP-52 is an active star which shows both out-of-transit photometric variability and starspot crossings during transits, we account for the contribution of non-occulted active regions in the retrieval. We recover a $0.1-10\times$ solar atmospheric composition, in agreement with core accretion predictions for giant planets, and a weak contribution of aerosols. We also obtain a $<3000$ K temperature for the starspots, a measure which is likely affected by the models used to fit instrumental effects in the transits, and a 5\% starspot fractional coverage, compatible with expectations for the host star's spectral type. Such constraints on the planetary atmosphere and on the activity of its host star will inform future \textit{JWST} GTO observations of this target.
\end{abstract}

\begin{keywords}
planets and satellites: atmospheres -- stars: starspots -- techniques: photometric -- techniques: spectroscopic
\end{keywords}



\section{Introduction}\label{intro}

Probing physical processes in exoplanet atmospheres that are thought to depend strongly on planetary mass, gravity and temperature requires a sample of targets spanning this phase space. WASP-52b was selected as a prime target for two large {\it Hubble Space Telescope} (\textit{HST}) programs (GO-14260, PI Deming and GO-14767, PIs Sing and L\'opez-Morales) because it occupies an important place in the continuum of mass, gravity and equilibrium temperature for hot Jupiters ($M\sim0.5\mathrm{M}_\mathrm{J}$, $\log g \sim2.9$, $T_{\mathrm{eq}}\sim1300~\mathrm{K}$). Given trends noted in the amplitude of water features \citep[e.g.][]{stevenson2016,fu2017}, WASP-52b is predicted to be on the boundary of hot Jupiters whose transmission spectra are strongly affected by the presence of clouds in the planet's atmosphere, and where large scale heights produce strong atmospheric signals. WASP-52b's mass lies in between that of Jupiter and Saturn's, which also suggests that it might be enriched in elements heavier than hydrogen and helium \citep[e.g.][]{thorngren2016}. The mass of this planet also falls in a range where atmospheric metallicity measurements are not available yet \citep[][and references therein]{wakeford2018}. Moreover, it is hosted by a relatively bright, mag$_\mathrm{H} = 10.1$ star. In many ways WASP-52b is an ideal target for exoplanet atmospheric characterization efforts, but one must carefully consider the influence of its moderately active K2V host star on observed spectra \citep{kirk2016,louden2017,mancini2017,chen2017_w52,alam2018,bruno2018_w52,may2018}.

Stellar activity is indeed one of the main challenges to correctly interpreting exoplanet observations. Starspots and faculae, in particular, are known to affect measured scattering slopes and molecular abundances \citep[e.g.][]{pont2007,lecavalierdesetangs2008,sing2011,ballerini2012,mccullough2014,rackham2017,pinhas2018,wakeford2019}. Such heterogeneities on the stellar photosphere have different effects whether they are occulted or not during a transit: non-occulted dark starspots and faculae produce an apparent, wavelength-dependent increase and decrease of the transit depth, respectively, and the apparent variation is opposite if the features are occulted \citep[e.g.][]{czesla2009,silva-valio2011,desert2011,bruno2016}. 
In specific spectroscopic channels, active regions can imprint stellar features on a transmission spectrum, especially in the case of cool, low-mass stars \citep[e.g.,][]{rackham2017,rackham2018}. Stellar contamination was shown to be less problematic for weakly-active FGK stars, but a case-by-case study is still warranted \citep{rackham2019}. When observations from different epochs are available, the problem of taking the changing stellar photosphere into account also arises \citep{barstow2015}, as measurements obtained in different epochs could otherwise be offset one from each other. When observations in different wavelengths and from different epochs are combined, both effects leave their trace on the planetary spectrum. 

The transmission spectra of WASP-52b have been observed separately with WFC3-IR \citep{bruno2018_w52} and with STIS and \textit{Spitzer}/IRAC \citep{alam2018}. In each case, stellar activity was taken into account with a different method. Here, we present a combined fit of the spectra with atmospheric retrievals, which we used to explore different ways of taking the time-varying impact of stellar activity into account. We describe the analysis carried out and the constraints obtained for both the planetary atmosphere and the activity features. In section \ref{obs_sect}, we describe the previously secured observations, and in section \ref{retrieval} detail how we implemented activity features in our retrieval scheme. In section \ref{results} and \ref{discussion} we present and discuss our results, respectively, and we conclude in section \ref{conclusions}.

\section{Observations}\label{obs_sect}
We obtained \textit{HST}/STIS (290-1030 nm) and \textit{Spitzer}/IRAC (3.6 and 4.5 $\mu$m) observations of WASP-52b in November 2016 (G430L grating), May 2017 (G750L grating), October 2016 (IRAC $3.6 \, \mu\mathrm{m}$ channel) and March 2018 (IRAC $4.5 \, \mu\mathrm{m}$ channel). The WFC3-IR observations (1.1-1.7 $\mu$m) were secured in August 2016 (G141 grism). Details on the acquisition, image processing and transit analysis were discussed in \citet{alam2018} for the STIS and IRAC data, and \citet{bruno2018_w52} for the WFC3 data.

In \citet{alam2018}, each spectroscopic transit was corrected for stellar activity by taking advantage of a semi-continuous, ground-based photometric monitoring of the host star. WASP-52 was observed with the All-Sky Automated Survey for Supernovae (\textit{ASAS-SN}) and with the Tennessee State University Celestron 14 (C14) Automated Imaging Telescope (AIT) at Fairborn Observatory (see Appendix A for details). The flux dimming, assumed to be caused by non-occulted dark starspots, was used to correct the time-series observations with \citet{sing2011} and \citet{huitson2013}'s prescriptions. In order to perform the correction, an effective temperature of 4750 K was assumed for the activity features.

\citet{bruno2018_w52}, on the other hand, observed a possible starspot crossing in the WFC3 transit. By using an analytic model for starspot occultations in transits \citep{montalto2014}, they fitted a starspot effective temperature of $\sim 4000$ K, which follows expectations for the host spectral type \citep{berdyugina2005}. However, no monitoring of the host star was available to the authors, so that no correction for non-occulted starspots was performed on the data set. In the hypothesis that dark starspots were dominant in the WFC3 observations as well, the baseline of the corresponding transmission spectrum was likely overestimated \citep[e.g.][]{czesla2009,sing2011}. In this work, we adopted the transmission spectrum resulting from their modeling of the starspot crossing.

\begin{figure}
\includegraphics[width=\columnwidth]{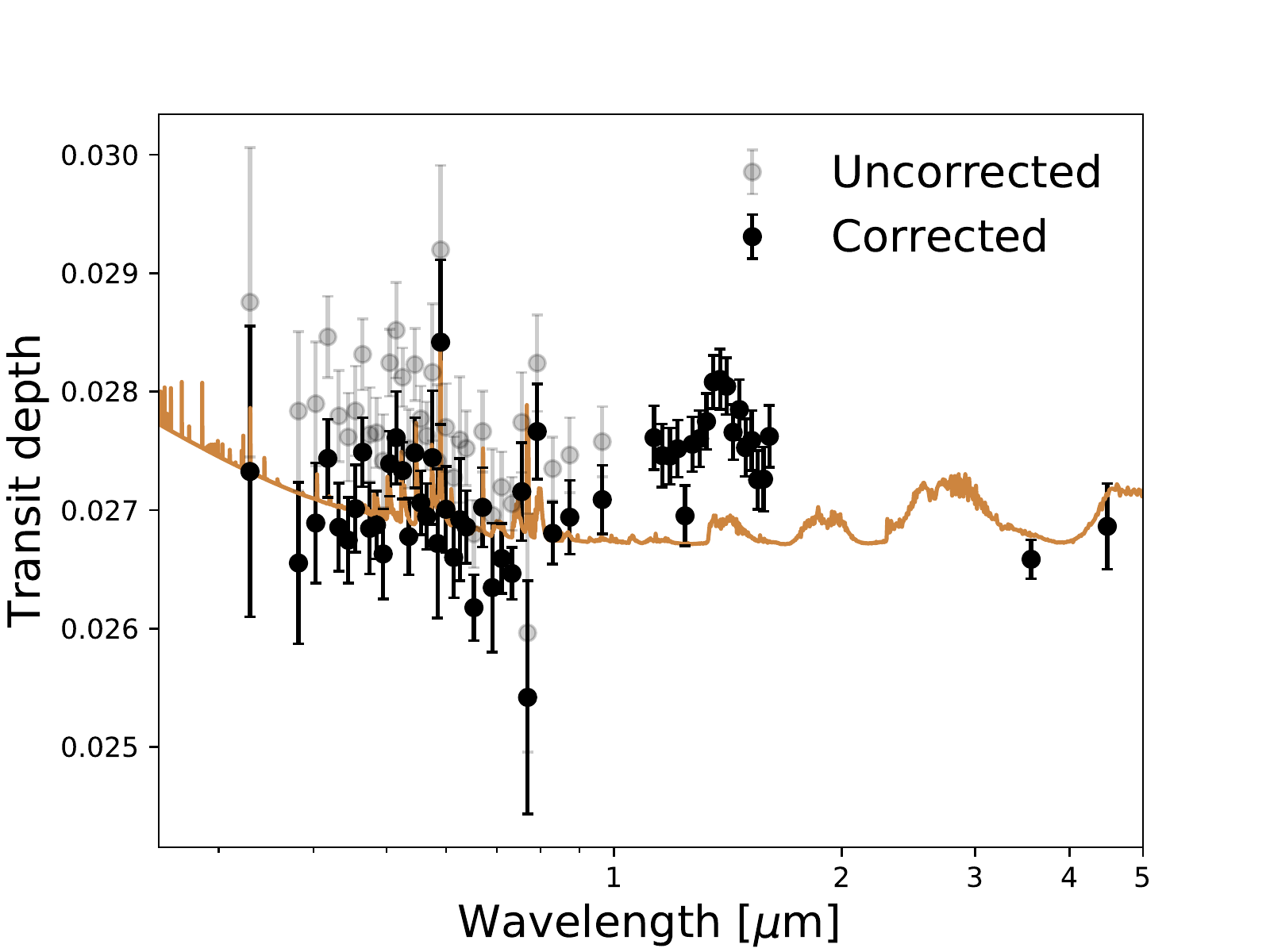}
\caption{\textit{HST}/STIS ($\simeq 0.3-1.0 \, \mu$m), WFC3-IR ($\simeq 1.1-1.7 \, \mu$m) and \textit{Spitzer}/IRAC (3.6 and $4.5 \, \mu$m) transmission spectra of WASP-52b, and the best fitting forward atmospheric model from \citealp{alam2018} (with equilibrium temperature \teq$=1315$ K, solar metallicity, C/O$=0.70$, a thick cloud deck and slight Rayleigh scattering slope). The STIS points before and after the correction for non-occulted starspots carried out by \citealp{alam2018} are shown in light gray and black, respectively. The STIS and WFC3 data sets are separated by a clear systematic offset. Due to the very small effect of unocculted starspots in the \textit{Spitzer} band, the respective uncorrected and corrected data points are indistinguishable.}
\label{datasets}
\end{figure}

In Figure \ref{datasets}, the STIS-WFC3-\textit{Spitzer} data sets are presented, both with and without the activity correction carried out by \citet{alam2018}. With such a wavelength coverage, an atmospheric retrieval exercise can in principle constrain the scattering properties of the atmosphere, its metallicity and cloud top pressure \citep{benneke2012, sing2016}, as well as allow inferences on the carbon-to-oxygen ratio thanks to the IRAC points in the infrared (IR). In order to combine the data sets into a consistent transmission spectrum, however, it is necessary to take the contribution of stellar activity into account: if the offsets between the observations in different epochs and wavelength windows are not corrected, the retrieval will try to compensate for the different baselines by changing parameters such as the molecular abundances or scattering factors, thereby biasing the result \citep{barstow2015,rackham2018}. 

\section{Retrieval scheme} \label{retrieval}

Retrieving the atmospheric properties of an exoplanet orbiting an active star requires several factors to be taken into account. Here, we studied the contribution of each aspect to the final result in a separate step of the analysis. First, we compared two methods of applying the starspot correction, as each one can potentially impact the transmission spectrum in a different way \citep[e.g.][]{rackham2018}. We alternatively parametrized the correction with the starspot distribution's average temperature or with the photospheric starspot fractional coverage, as we discuss in Section \ref{sectspot}. Then, we stitched our observations together by correcting either the visible or the IR part of the spectrum, as we detail in Section \ref{stitch}. Indeed, as our observations were obtained in different epochs, they were affected by changing starspot configurations, in spectral regions that are not equally sensitive to stellar activity. Finally, we compared the results of two of the several atmospheric retrieval codes that are currently available. Such codes can either assume chemical equilibrium in the planet atmosphere or a setting where molecular abundances are free parameters. We present the atmospheric models and codes used for the retrievals in Section \ref{frameworks}.

\begin{figure*}
\includegraphics[width=0.8\textwidth]{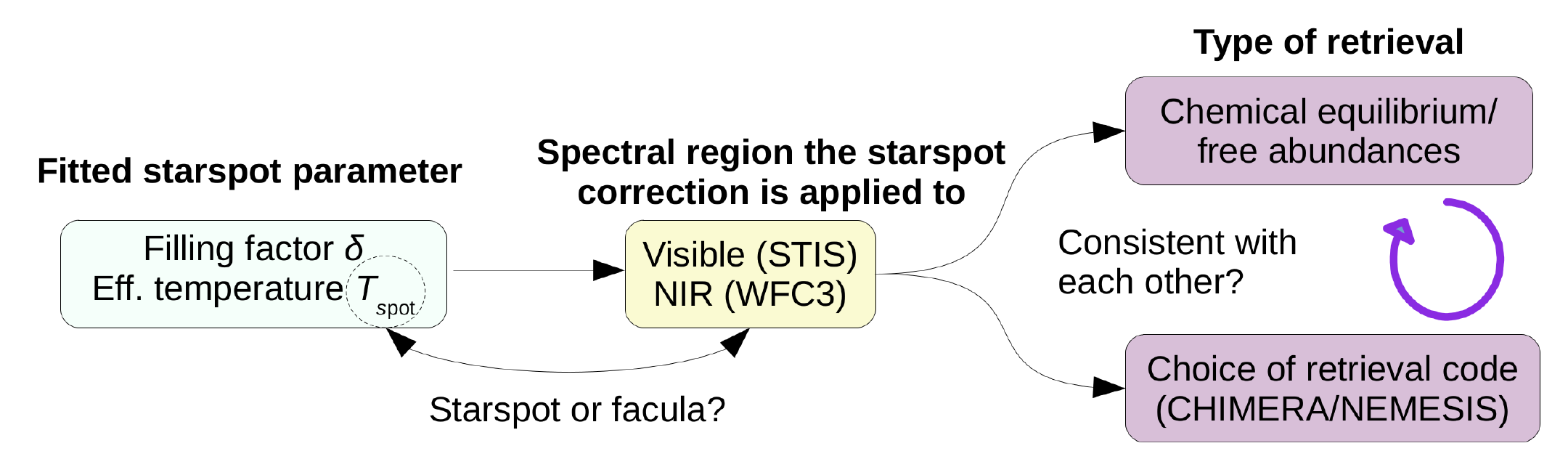}
\caption{Outline of the retrieval analysis. First, the type of starspot parameter to fit for is chosen (Setion \ref{sectspot}). Then, the spectral window on which to carry the starspot correction out is selected (Section \ref{stitch}); while the STIS data require a facula in order for the visible spectrum to be stitched to the IR data, the opposite is true for the WFC3 measurements. Finally, the atmospheric model is used either in the chemical equilibrium or the free abundance mode, the latter being tested with two different retrieval codes (Section \ref{frameworks}). Results obtained with different approaches should be in agreement with each other.}
\label{flowchart}
\end{figure*}

An outline of our analysis, which involved numerous iterations between each step in order to reach an optimal solution, is presented in Figure \ref{flowchart}.

\subsection{Adding starspots to the retrieval}\label{sectspot}
Non-occulted starspots can be modeled by using synthetic stellar spectra with a different effective temperature than the star. The resulting stellar spectrum can then be computed as a combination of the ``nominal'' stellar spectrum and a fractional component of such a lower temperature spectrum.\footnote{A more precise formulation would be the use of a lower-$\log g$ stellar model, because the magnetic pressure within a starspot reduces the gravitational acceleration at that point of the stellar photosphere. This can be modeled with a lower-$\log g$ stellar model (J. Valenti, S. H. Saar, priv. comm.), but to our knowledge this aspect has not yet been quantitatively explored in the literature. Therefore, we used the simplifying assumption of having starspots with the same $\log g$ as the star.} The contribution of starspots to the total stellar flux increases as their brightness contrast with the stellar surface $F_\mathrm{spot}/F_\star$, which is a function of the difference between their effective temperature $T_\mathrm{spot}$ and the stellar effective tempereature $T_\star$, decreases. A non-occulted starspot which is cooler than the average stellar photosphere will produce a wavelength-dependent apparent increase in the transit depth, and conversely a non-occulted facula, which is warmer than the stellar photosphere, will produce shallower transits.\footnote{This is another approximation, as faculae are not described by the same limb darkening law of starspots, and are actually brighter at the stellar limb \citep[e.g.][and references theirein]{norris2017}.} 
As a result, the average transit depth corresponding to a given epoch will be shifted compared to the ``true'' depth by an amount that depends on the starspot fractional coverage (or filling factor $\delta$) and effective temperature $T_\mathrm{spot}$. 

In our analysis, we modeled the integrated effect of non-occulted starspots on the stellar spectrum by combining stellar and starspot synthetic models. With this approach, we disregarded whether the star was covered by a single, giant feature, or by a group of small features. We set the starspot effective temperature $T_\mathrm{spot}$ and filling factor $\delta$ as free parameters, but fit only one of them at a time. Without additional constraints, these parameters are indeed at least partially degenerate for non-occulted activity features, contrary to high signal-to-noise observations of starspot occultations \citep[e.g.][]{sing2011}. This allowed us to test whether different modeling approaches, as well as different ways of taking the constraints from photometry into account, produce results in agreement for our target.

An example of how to add the contribution of a heterogeneous stellar surface to the transmission spectrum can be found in \citet{rackham2017}. These authors explored both composition (metallicity) and temperature heterogeneities in the stellar photosphere. We instead decided to explore the contribution of a stellar photosphere with only temperature heterogeneities, as these can more directly represent dark and bright active regions which can cause transit depth offsets.

\begin{figure*}
\includegraphics[width=\textwidth]{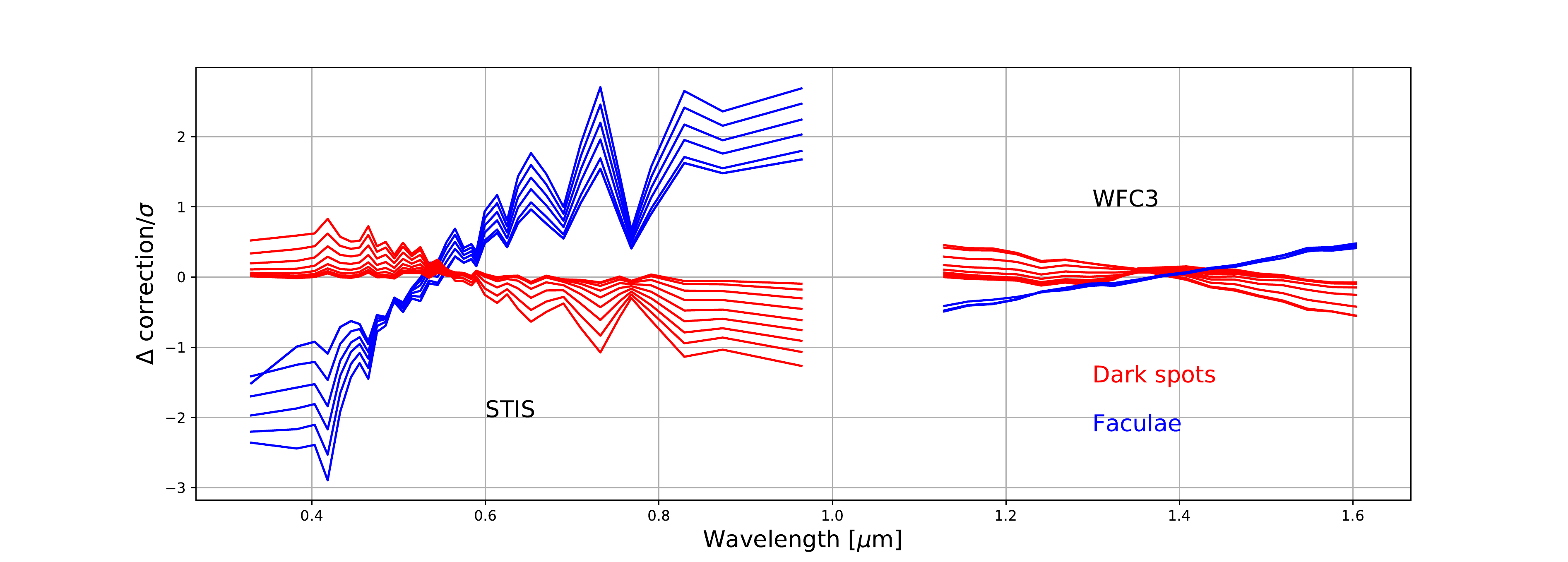}
\caption{Wavelength-dependent correction variation with respect to the whole filter correction ($D_\lambda - D_{\lambda, 0}$) across the different STIS and WFC3-IR channels, divided by the uncertainty of the transmission spectrum measurements. The red lines show the correction obtained for dark spots (from 2300 to 4700 K in 300 K steps), and the blue ones the correction obtained for faculae (from 5100 to 6600 K in 300 K steps). This plot shows that the impact of non-occulted activity features on different channels is non-negligible in the visible, while it is smaller in the near-IR. The apparent discontinuity between the STIS and the WFC3 correction is due to the multiplying factor $D_{\lambda, 0}$ (equations \ref{delta} and \ref{ts}).}
\label{channels}
\end{figure*}

Starspots and faculae have a time-dependent effect on the transit depth, according to the amount of active features that cross the stellar visible disk as the star rotates. Each feature has its own spectrum, and therefore a wavelength-dependent effect on the transit depth. Either the starspot fractional coverage $\delta$ or the average starspot effective temperature $T_\mathrm{spot}$ can be used to compute the transit depth variation at each epoch. These two parameters can then be deduced with an atmospheric retrieval.

When retrieving for the starspot fractional coverage $\delta$, we used the formula \citep[][and references therein]{rackham2017}
\begin{equation}
    \frac{D_\lambda - D_{\lambda, 0}}{D_{\lambda, 0}} = \frac{1}{{1 - \delta (1 - F_{\lambda, \mathrm{spot}}/F
    _{\lambda, \star})}} - 1,
\label{delta}
\end{equation}
where $ D_\lambda$ is the observed transit depth, $D_{\lambda, 0}$ the true depth, and $F_{\lambda, \mathrm{spot}}$  and $F_{\lambda, \star}$ are the starspot and stellar brightness, respectively. In this case, we relied on the starspot temperature for $\sim 5000$~K stars provided by \cite{berdyugina2005}, i.e. $\sim 3700$~K. When fitting for the starspot effective temperature $T_\mathrm{spot}$, we referred to \citet{sing2011}'s formula,
\begin{equation}
\frac{D_\lambda - D_{\lambda, 0}}{D_{\lambda, 0}} = \Delta f(\lambda_0)\frac{1 - F_{\lambda, \mathrm{spot}}(T_\mathrm{spot})/F_{\lambda, \star}}{1 - F_{\lambda_0, \mathrm{spot}}(T_\mathrm{spot})/F_{\lambda_0, \star}},
\label{ts}
\end{equation}
where $\Delta f(\lambda_0)$ is a factor depending on the photometric monitoring of the star, observed at $\lambda_0$, and is a measure of the stellar flux dimming due to non-occulted starspots. It is therefore related to the starspot filling factor. To determine it, we adopted the Gaussian process (GP) computed by \citet{alam2018} on the \textit{ASAS-SN} and \textit{AIT} data. We calculated $\Delta f(\lambda_0) = 1 - f_\mathrm{norm} = 0.031$, where $f_\mathrm{norm}$ is the mean of the GP prediction normalized to the non-spotted stellar flux for the whole observation window. As explained by \citet{alam2018}, such non-spotted stellar flux was calculated as $F_\star = \max (F) + k \sigma$, where $F$ is the \textit{ASAS-SN} and \textit{AIT} variability monitoring data, $\sigma$ is the dispersion of the photometric measurements and $k$ is a factor fixed to unity \citep{aigrain2012}.

In both cases, $F_{\lambda, \mathrm{spot}}$  and $F_{\lambda, \star}$ were evaluated with PHOENIX stellar models \citep{husser2013}, integrated for every spectroscopic channel across the respective instrument filter.\footnote{The filters were downloaded from the \texttt{svo} filter profile service, \texttt{http://svo2.cab.inta-csic.es/theory/fps3/.}} To reduce the computation time of the retrieval (as discussed below), we prepared a grid containing the correction factors for a range of starspot temperatures, from 2300~K to 6000~K in 100~K intervals, as well as for different filling factors, from 0.1\% to 30\% in steps of 1\%. At each iteration of the retrievals, the correction was evaluated with a bilinear interpolation between the model corresponding to closest cooler (or lower $\delta$) model and closest warmer (or higher $\delta$) model in the grid by using the \texttt{Python} function \texttt{interp2d} \citep{scipy}.

These two formulations (fit for starspot temperature and filling factor) allowed us to test different approaches, and explore whether they drive the retrieval to the same solutions. With equation \ref{delta}, we ignored the constraints from the photometric monitoring and retrieved the atmospheric status by assuming a $T_\mathrm{spot}$ value as those reported in the literature for K stars. With equation \ref{ts}, we used our knowledge of the out-of-transit light curve behaviour and derived $T_\mathrm{spot}$ without setting priors on the most likely filling factor. Such two methods can potentially lead to different results in certain bands, insofar as a small, particularly cold starspot has a different contribution than a large, warmer one (or of a large group of small spots, as shown by \citealp{rackham2018}).

In order to limit the number of free parameters, we fitted for configurations where a starspot (or group thereof) would affect only one data set at a time. We could have simultaneously modeled a starspot for each data set, but the effects of multiple starspots would have canceled out. Moreover, little information would have been available in order to break the starspots' degenerate effects in different epochs.

\subsection{Stitching the full optical to IR spectrum together}\label{stitch}
In photometry, a few percent variation of WASP-52's brightness can be observed across few stellar rotations \citep[e.g.][]{affer2012,basri2013}. As the ground-based monitoring with \textit{ASAS-SN} and \textit{AIT} showed \citep{alam2018}, all our observations are affected by stellar activity features by a different amount. While the optical spectrum was corrected for stellar activity by relying on photometric monitoring and assumptions on the starspot temperature \citep{alam2018}, no correction for non-occulted starspots was adopted by \cite{bruno2018_w52}.

As motivated in Section \ref{obs_sect}, both corrections are affected by a certain degree of uncertainty. In the visible, the photometric monitoring does not constrain the starspot temperature, and an e.g.  excessive correction would imply an underestimated transit depth. In the IR, the lack of constraints on non-occulted starspots might imply an overestimated transit depth. The need of stitching the optical and IR component into a single, consistent transmission spectrum can however provide constraints on the possible bias affecting both data sets. The different epochs of observation would require at least two activity features to be included in the retrieval, one for the optical part of the spectrum and another for the IR. However, given the likely opposite corrections needed (increased transit depth in the optical, decreased in the IR), the effect of the two features would likely be degenerate. Hence, we decided to explore the two scenarios separately, noticing that this might determine only upper or lower limits on the starspot or faculae parameters, instead of precise constraints. 
In the first, we used the correction for stellar activity derived by \citet{alam2018} on the STIS points, and retrieved the starspot parameters on the WFC3 data set. In the second, we disregarded the photometric constraints and shifted the STIS data set in order for it to match WFC3 and \textit{Spitzer} spectra. In this latter case, we assumed no need to correct the WFC3 data, as if the effect of stellar activity in the IR were negligible (as it is often assumed) and so measurements in this spectral window could be used as an ``anchor''. We verified that the \textit{Spitzer} points were not sensitive to the starspot correction, contrary to the WFC3 points.

We remark that the WFC3 spectrum could have been corrected by a single, G141-filter-wide value, without the need to calculate the correction factor for every channel. We demonstrate this in Figure \ref{channels}, where we plot the difference between the correction factor for every channel and the mean correction across all channels, divided by the uncertainty on the data points. In other words, we show the result of integrating the starspot and stellar spectra across every channel instead of integrating them only once for the whole spectrum, and weight such difference with the uncertainty of the observations. The plot shows the correction to be applied both for dark starspots and for faculae. Because of the stronger signal of starspots in the optical compared to the IR, the correction can vary by $\pm 3 \sigma$ compared to the average across the STIS channels, while the correction for WFC3-IR channels only changes by about $\pm 0.4 \sigma$. The integration time for the retrieval increases as the number of channel integrations increases,  and because of this we considered pre-computing the correction grid as described in Section \ref{sectspot}. This also enabled us to retrieve the correction of the WFC3 data for every channel with a relatively small computational effort, even if, as just shown, this was not strictly necessary. 

\subsection{Comparing retrieval frameworks}\label{frameworks}
A variety of atmospheric retrieval approaches are currently available \citep[see e.g. the review by][]{madhusudhan2018}, but few comparisons between different methodologies and codes have been carried out \citep[e.g.][]{kilpatrick2018,barstow2018}. We therefore compared two retrieval codes, described below. In the first case, we implemented a starspot correction in the retrieval scheme, where stellar models were used at each iteration to ``shift'' the output of the atmospheric model. In the second case, we used the best solution of the first approach (with a reduced chi-square $\tilde{\chi}^2 \simeq 1$) to correct the observed transmission spectrum and fed it to another retrieval code. The first code also allowed us to compare both a retrieval with and without adopting chemical equilibrium.

\subsubsection{Retrievals with \texttt{CHIMERA}}
The \texttt{CHIMERA} software \citep{line2013,kreidberg2014} can generate a transmission spectrum given a pressure-temperature (\textit{P-T}) profile for the planet atmosphere (described by irradiation temperature\footnote{The irradiation temperature is defined as the temperature at the substellar point of an exoplanet \citep[see e.g.][]{heng2017book}.} $T_{\mathrm{irr}}$, IR opacity $\kappa_{\mathrm{IR}}$ and ratio of visible to IR opacity $\gamma_1$), global abundances (metallicity with respect to solar $[\mathrm{M/H}]$ and carbon-to-oxygen abundance ratio $\mathrm{C/O}$), carbon and nitrogen quenching pressures\footnote{The quench pressure is the pressure below which (and corresponds to the altitude above which) the mole fraction of a species remains constant. This usually happens when the temperature of a system drops low enough for the transport time scale to become shorter than the chemical time scale required to maintain that species in equilibrium with other constituents; pressure changes can also be involved \citep{prinn1977}}  ($P_\mathrm{q}(C)$ and $P_\mathrm{q}(N)$), scattering cross section and slope ($\sigma_0$ and $\gamma$, respectively), grey cloud-top pressure ($P_{\mathrm{c}}$) and scaling factor for the planetary radius at 10 bar ($xR_p$), which takes the uncertainty on the planetary radius into account. The molecular abundances at chemical equilibrium are calculated from elemental abundances with the NASA \texttt{CEA2} code.\footnote{\texttt{https://www.grc.nasa.gov/WWW/CEAWeb/ceaHome.htm}.} \texttt{CHIMERA} uses \texttt{PyMultiNest}, a software for Bayesian inference through multimodal nested sampling \citep{feroz2008,feroz2009,buchner2014}. We added the two starspot parameters ($T_\mathrm{spot}$ and $\delta$) as free parameters in the retrieval, and explored their posterior distributions together with those of the atmospheric model. 

Because of the lack of constraints on the atmospheric \textit{P-T} profile coming from transmission spectra, we modeled an isothermal profile by fixing $\kappa_{\mathrm{IR}}$ to a low value ($\simeq 0.03 \, \mathrm{cm}^2\mathrm{g}^{-1}$) and setting visible opacity equal to near-IR (NIR) opacity ($\gamma_1 = 1$). We also neglected quenching phenomena by fixing both $P_\mathrm{q}(C)$ and $P_\mathrm{q}(N)$ to $1 \, \mu\mathrm{bar}$ (i.e., confining quenching at lower pressures than those probed by transmission spectroscopy). \texttt{CHIMERA} also allows the modeling of transmission spectra produced in out-of-chemical equilibrium configurations, where the abundances of individual molecules are free parameters. 
We explored this retrieval mode and recovered the abundances of water, methane, carbon monoxide, sodium and potassium while fixing hydrogen and helium to their solar value and deriving molecular hydrogen in order to obtain a total atmospheric abundance equal to unity.

\subsubsection{Retrievals with \texttt{NEMESIS}}
\texttt{NEMESIS} couples a fast correlated-k \citep{lacis1991} forward model with \texttt{PyMultiNest}, and has been extensively used to model both Solar System objects and exoplanets. \citet{barstow2017} performed a retrieval study exploring cloudy solutions for transmission spectra of 10 hot Jupiters \citep{sing2016}. With the recent inclusion of the \texttt{MultiNest} algorithm, improvements to the cloud retrieval scheme used in \citet{barstow2017} have been made \citep{krissansentotton2018}. Instead of a grid search approach covering a range of pre-defined cloud top and base pressures, and exploring only Rayleigh or grey clouds, the cloud top and base pressures and index of the scattering slope are now free parameters within the model. The cloud is thus represented by these last three parameters and by the total optical depth. Other model parameters include the abundances of H$_2$O, CO, CH$_4$, Na and K, an isothermal temperature and the radius at the 10 bar pressure level. Atomic absorption data comes from the VALD database \citep{heiter2008}. H$_2$--H$_2$ and H$_2$--He collision-induced absorptions are taken from \citet{borysowfr1989,borysowfr1990}, \citet{borysowetal1989,borysow1997} and \citet{borysow2002}.

For the retrieval with \texttt{NEMESIS}, we did not include any starspot in the model, i.e. we did not fit for the offset between data sets, but used the stellar activity-corrected WFC3 data set obtained from the best solution of the \texttt{CHIMERA} chemical-equilibrium retrieval. We relied on the activity-corrected STIS and \textit{Spitzer} data obtained by \cite{alam2018} for the rest of the transmission spectrum.

\section{Results}\label{results}

\subsection{Starspot temperature vs. filling factor}\label{chemeqret}
\begin{figure*}
\includegraphics[width=0.8\textwidth]{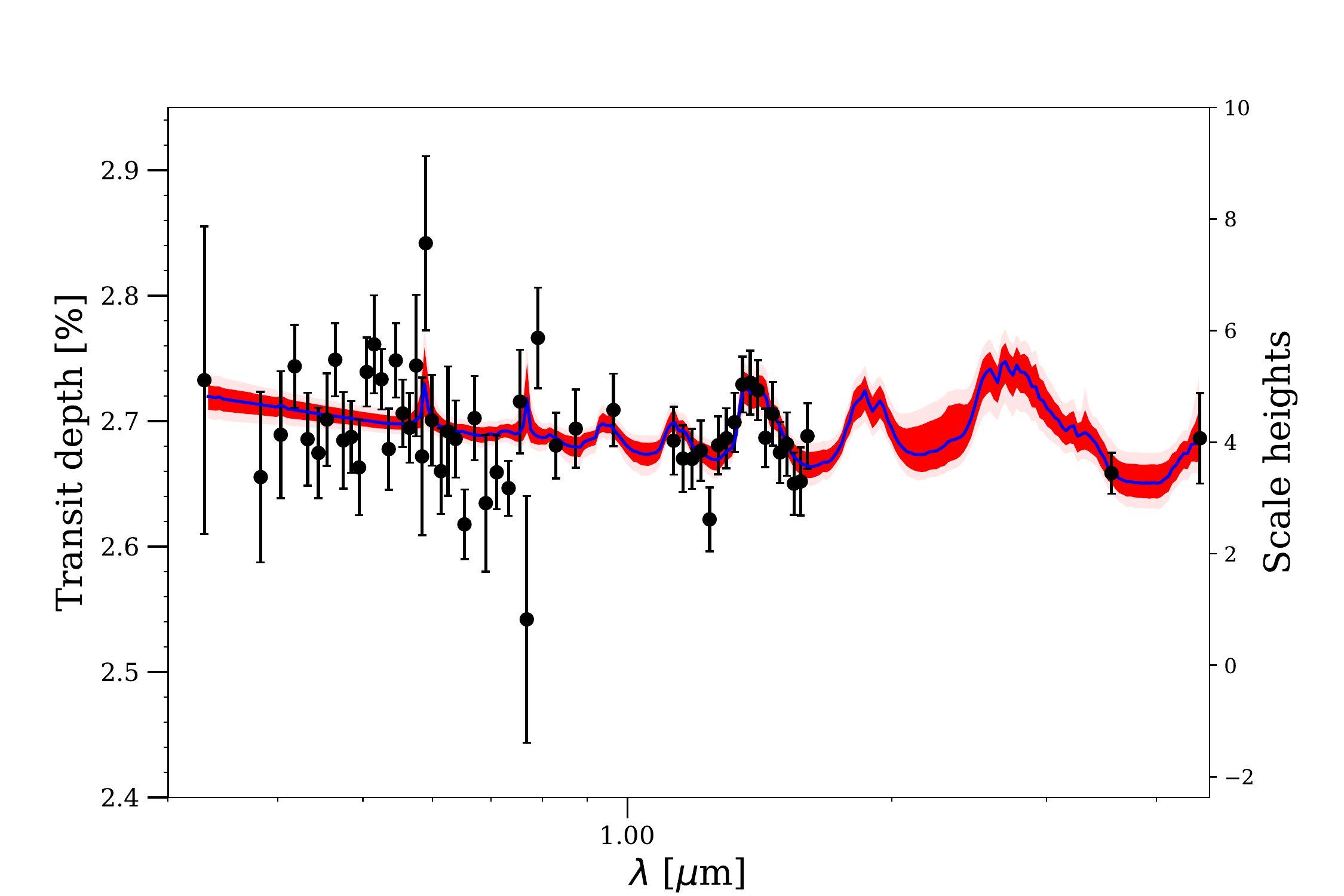}
\caption{Best solution (blue curve) and 1 and $2\sigma$ credible intervals (red) for the \texttt{CHIMERA} chemical-equilibrium retrieval where the starspot parameters are derived from the WFC3 data set. The WFC3 data has been shifted with Eq. \ref{delta}, due to the starspot correction.}
\label{Tspot_spec}
\end{figure*}

We first describe the results of the retrievals with \texttt{CHIMERA} in chemical equilibrium, where we used the STIS data as corrected by \cite{alam2018} and added an activity feature acting on the WFC3 data. Figure \ref{Tspot_spec} presents the best fit and the 1 and $2\sigma$ credible intervals of the retrieval with varying $T_\mathrm{spot}$, where we observe that the average transit depth of the WFC3 points has been reduced -- which indicates a dark starspot acting on the spectrum. The best-fit solution obtained by the retrieval corresponds to a reduced chi-square of $\tilde{\chi}^2 = 1.15$, as indicated in Table \ref{tabres} together with the 50th and 68.2th percentiles for each retrieved parameter. The marginalized posterior distributions are presented in the orange contours of Figure \ref{Tspot_corner}. We recovered a $\sim 0.1-10\times$ solar metallicity, subsolar C/O and a $\simeq 2770^{+430}_{-300}$~K starspot. The broad wavelength coverage of our data set allowed the retrieval to break the degeneracy between $T_\mathrm{spot}$ (or $\delta$) and the scattering parameters $\gamma$ and $\sigma_0$. This is due, in particular, to the different action of starspots in the visible (where this correlation is stronger, \citealp[e.g.][]{sing2011}) and in the IR. The cloud top pressure goes from $\sim10$ mbar up to $\sim 100$ bar, indicating that both solutions with and without clouds are compatible with the observed transmission spectrum (which only probes 1 to 100 mbar atmospheric features). 
Partial muting of the spectral features (in particular, water at $1.4 \mu$m) was obtained by the retrieval with the use of high [M/H] or low scattering cross-section, which are indeed correlated.

\begin{table*}
\caption{Results of the \texttt{CHIMERA} retrievals with an active region correction applied to the WFC3-IR data set, and of the \texttt{NEMESIS} retrievals on the corrected optical to IR transmission spectrum. Columns separate the retrievals with chemical equilibrium (CE) and without (No CE). The ``method'' row indicates whether the starspot effective temperature $T_\mathrm{spot}$, filling factor $\delta$ or none was fitted.}
\label{tabres}
\begin{center}
\begin{tabular}{l|rr|rr}
\hline \hline
 & \multicolumn{2}{c}{CE (\texttt{CHIMERA})} & No CE (\texttt{CHIMERA}) & No CE (\texttt{NEMESIS}) \\
 \hline
Method & $T_s$ & $\delta$ & $T_s$ & None\\
\hline
$T_{\mathrm{irr}}$ [K] & $1020^{+217}_{-185}$ & $961^{+220}_{-184}$ & $579^{+187}_{-118}$ & $625^{+130}_{-121}$\\
$\log [\mathrm{M/H}]$ & $-0.19^{+0.93}_{-1.09}$ & $-0.35^{+1.05}_{-1.42}$ & $0.12^{+0.94}_{-1.12}$ & $-1.25^{+0.44}_{-0.41}$\\
$\log (\mathrm{C/O})$ & $-1.27^{+0.59}_{-0.47}$ & $-1.30^{+0.64}_{-0.47}$ & -- & --\\
$\log \mathrm{H}_2\mathrm{O}$ & $-3.30^{+1.07}_{-0.99}$ & $-3.24^{+1.05}_{-1.51}$ & $-3.30^{+0.94}_{-1.12}$ & $-4.67^{+0.44}_{-0.41}$\\ 
$\log \mathrm{CH}_4$ & $-6.85^{+2.12}_{-2.15}$ & $-6.26^{+2.10}_{-2.26}$ & $-9.17^{+1.89}_{-1.78}$ & $-9.51^{+1.68}_{-1.57}$\\
$\log \mathrm{CO}$ & $-4.62^{+1.18}_{-1.20}$ & $-4.97^{+1.42}_{-1.87}$ & $-7.25^{+3.10}_{-3.04}$ & $-7.62^{+2.69}_{-2.75}$ \\
$\log \mathrm{Na}$ & $-6.14^{+0.94}_{-1.05}$ & $-6.75^{+1.25}_{-1.37}$ & $-8.33^{+2.28}_{-2.18}$ & $-7.04^{+1.97}_{-3.07}$ \\
$\log \mathrm{K}$ & $-7.14^{+0.98}_{-0.99}$ & $-7.43^{+1.12}_{-1.25}$ & $-8.42^{+2.33}_{-2.15}$ & $-7.84^{+1.96}_{-2.59}$\\
$\sigma \, [\log \sigma_{\mathrm{H}_2}]$ & $1.56^{+0.82}_{-0.97}$ & $1.40^{+0.96}_{-1.13}$ &  $0.45^{+1.01}_{-0.86}$ & -- \\
$\gamma$  & $1.24^{+0.69}_{-0.46}$ & $1.75^{+1.08}_{-0.73}$ & $1.82^{+1.09}_{-0.77}$ & $7.23^{+4.34}_{-4.61}$\\
$\log P_c$ (cloud top pressure, bar) & $0.20^{+1.74}_{-1.75}$ & $0.28^{+1.70}_{-1.72}$ & $0.28^{+1.66}_{-1.70}$ & $-4.92^{+3.05}_{-2.04}$\\
$xR_p$ & $0.97\pm0.01$ & $0.97\pm0.01$ & $0.99^{+0.01}_{-0.02}$ & $0.99\pm0.00$\\
$\log$ cloud base pressure [bar] & -- & -- & -- & $-1.24^{+1.62}_{-2.57}$\\
$\log$ opacity [dimensionless] & -- & -- & -- & $-5.52^{+2.95}_{-2.90}$\\
$T_\mathrm{spot}$ [K] & $2770^{+434}_{-302}$ & -- & $2837^{+476}_{-347}$ & --\\
$\delta$ [\%] & -- & $5\pm1$ & -- & --\\
$\tilde{\chi}^2$ & 1.15 & 1.13 & 1.12 & 1.28\\
\hline 
\end{tabular}
\end{center}
\end{table*}

This retrieval required an unusually cool starspot for a K dwarf, which could mean that the STIS data was overcorrected by \citet{alam2018}. As these authors showed in their Figure 2, a \textit{warmer} starspot in the optical corresponds to a \textit{larger} contamination; to adopt the largest possible correction for their data, they used a 4750 K starspot. The opposite is true for the IR: as a result, the WFC3 data might need to be shifted ``too much'' to be compatible with the visible spectrum, and require too cool of a starspot for a K star. On the other hand, the occulted starspot correction carried out by \citet{bruno2018_w52} should not significantly affect the pre-corrected WFC3 data, as the spectroscopic transit depth they obtained from the fit of the occulted starspot (Table 5 of their paper) is, in most cases, a few hundreds ppm \textit{smaller} than the transit depth obtained without taking the starspot into account. This means that their occulted spot model requires a \textit{warmer} starspot than their spot-free case in order for the IR data to match the optical data. 

Another possibility is that the actual value of the GP model used to calculate $\Delta f(\lambda_0)$ (Section \ref{obs_sect}) was underestimated, so that a warmer starspot would be sufficient to achieve the same correction factor. This could be the case if the actual stellar flux at the WFC3 epoch was much different from its average value computed with the GP, but the photometric scatter makes this possibility hard to test. A third hypothesis is that the significant offset between the STIS and the WFC3 data sets is due to the different systematic models used in \citet{alam2018} and \citet{bruno2018_w52}, which might impact transit normalization and therefore the reference radius for the planet in the transmission spectrum. We can say, at the very least, that the retrieval with a non-occulted starspot acting on the WFC3 data set constrained a lower temperature limit for such a starspot (or for a group thereof). This might suggest that, if observations are taken in different epochs, the most convenient approach is to fit all data sets with the same system parameters, without correcting for stellar activity beforehand, and then correct for stellar activity on all data sets together.

The retrieval where we corrected the STIS data set (and therefore ignored \citealp{alam2018}'s correction) helped us to study these possibilities. In this case, we fitted a facula with $T_s \simeq 5040^{+20}_{-15}$ K, as well as a larger scattering feature ($\gamma \simeq 2.4\pm0.7$), which mutes the $1.4\, \mu$m water absorption (we show the result in Figure \ref{stisretrieval}). Because of this and of the resulting $\tilde{\chi}^2$ increment compared to the first case (1.88 against 1.15), we decided to reject this scenario. The planet atmospheric features provide therefore some constraints on the stellar features, as found in other cases in the literature \citep[e.g.][]{rackham2017, wakeford2019}.

We now turn to the case where the starspot fractional coverage of the stellar surface $\delta$ was retrieved from the WFC3 data. In this case, we adopted a starspot temperature $T_\mathrm{spot} = 3700 \, \mathrm{K}$, compatible with a $\sim 5000 \, \mathrm{K}$ dwarf star \citep{berdyugina2005}. Because shifting the STIS data set produces a model hardly compatible with WFC3 water absorption, we only consider the case where we assumed \citet{alam2018}'s correction on the STIS observations, and a non-occulted active region was fitted on the WFC3 data set. The blue contours in Figure \ref{Tspot_corner} represent the marginalized posterior distributions and 1 to $3\sigma$ credible intervals, and the 50th to 68.2th percentiles of the retrieved parameters are presented in Table \ref{tabres}. Also in this case, we found a good fit ($\tilde{\chi}^2 = 1.13$) and consistent results with the previous retrieval (shift of the WFC3 data set with $T_\mathrm{spot}$). For a filling factor $\delta = 5 \pm 1\%$, which denotes a weakly active K dwarf \citep{berdyugina2005}, we obtained again a $\sim0.1-10 \times$ solar metallicity and subsolar C/O, moderate scattering and a wide range of cloudy to clear atmospheres.

\subsection{Free abundances with \texttt{CHIMERA}}\label{freechimera}
In Table \ref{tabres}, we present the abundances of H$_2$O, CH$_4$, CO, Na and K extracted from the retrieval with \texttt{CHIMERA}-chemical equilibrium (Section \ref{chemeqret}), and compare them with the retrieval with free abundances (third column). The atmospheric statuses obtained without the assumption of chemical equilibrium are at $1-2\sigma$ agreement with those obtained when assuming it. In particular, we observed that the water abundance is compatible with the current knowledge of the solar composition, i.e. $\log (\mathrm{O/H})_\odot = -3.31 \pm 0.05$ \citep{asplund2009}. For this retrieval, the very low (0.1-100 parts per billion) carbon abundance is in agreement with the sub-solar C/O found by the retrieval with chemical equilibrium. However, the very large uncertainties, due to the absence of carbon species absorption features in the observed spectral range, prevent us from drawing any conclusion. 

We derived the metallicity from the free abundances by using the approach followed by \cite{kreidberg2014} and \cite{louden2017} for WASP-52b. Since WASP-52b and HD 189733b have similar equilibrium temperatures, the solar metallicity disequilibrium abundance calculations presented in \cite{moses2011} for the latter planet may be applied to this case. The terminator volume mixing ratio of H$_2$O is taken to be 383 parts per million per unit volume, and by comparing this to our retrieved values with \texttt{CHIMERA} we infer that the $1\sigma$ metallicity range is $0.1-11.5 \times$ solar (Table \ref{tabres}). As the only gas with a reliably constrained abundance is water, however, any estimate of metallicity based on this must be viewed with extreme caution. We also remark the difference in the \tirr~ recovered by the free-abundance and the chemical equilibrium retrieval ($\sim 600 \pm 150$ K and $\sim 1000 \pm 200$ K, respectively), which can be explained by the lack of constraints on this parameter offered by transmission spectra.

\citet{alam2018} reported a $2.3\sigma$ detection of the NaI doublet at 589.3 nm and a possible muting of the K absorption feature at 766.5 nm by clouds. In the results of our retrievals in Figure \ref{Tspot_spec}, such features are present both in the best models and in the $2\sigma$ credible intervals. However, both in the chemical equilibrium and in the free abundance case, we do not find significant constraints on the abundances of Na and K (Table \ref{tabres}), as demonstrated by the large uncertainties on their posterior distributions. 

\begin{figure*}
\includegraphics[width=\textwidth]{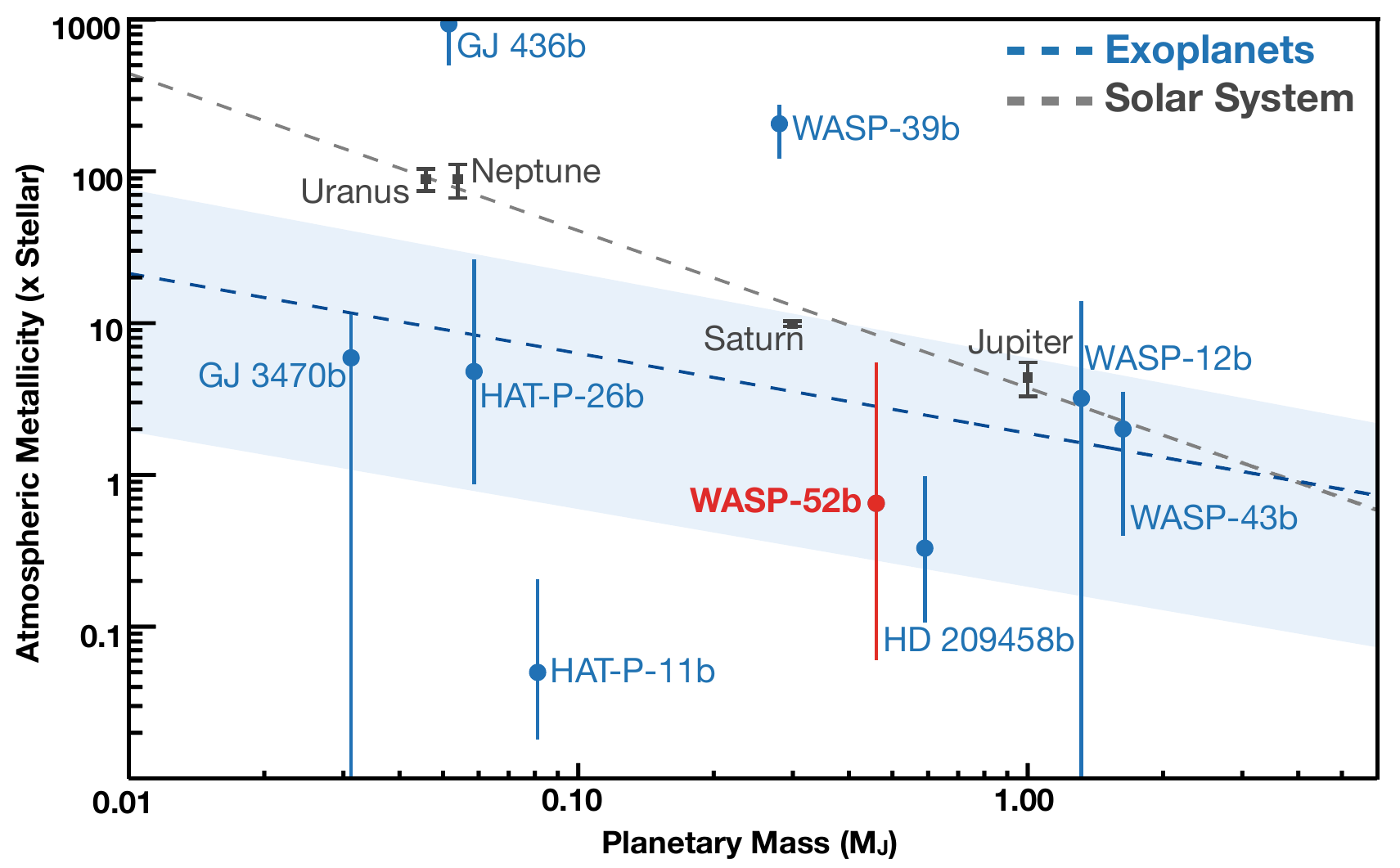}
\caption{Correlation between the mass and the atmospheric metallicity (i.e., the logarithmic fraction of elements heavier than helium, with respect to hydrogen) of giant planets. WASP-52b is shown in red. The black dashed line is the fit of the correlation for Solar System planets, and the blue dashed line and the light blue region represent the fit and its uncertainty for exoplanets with measured metallicity. Adapted from \protect\cite{wakeford2017}, with updates from \protect\cite{morley2017}, \protect\cite{benneke2019} and \protect\cite{chachan2019}.}
\label{massmet}
\end{figure*}

\subsection{Free abundances with \texttt{NEMESIS}}
The posterior distributions of the \texttt{NEMESIS} retrieval are shown in Figure \ref{nemesis}. The resulting molecular abundances are within $2 \sigma$ or less from the free-abundances \texttt{CHIMERA} retrieval. In particular, we observe that \texttt{NEMESIS} achieved a best-fit model with $\tilde{\chi}^2 = 1.28$, but using a $\sim 20 \times$ lower water abundance, with $2-3\times$ smaller uncertainties than the \texttt{CHIMERA} retrieval for water and methane. 

We derived the atmospheric metallicity from the free abundances following \cite{louden2017} and \cite{moses2011}, as done for the \texttt{CHIMERA} free abundances case. The calculation was carried out in order to facilitate comparison with the \texttt{CHIMERA} result, even if with the same caveats described earlier. Hence, from the retrieval with \texttt{NEMESIS}, we infer that the 1$\sigma$ metallicity range is $0.022 - 0.15 \times$ solar (Table \ref{tabres}, fourth column).

This latter metallicity has to be compared with the one calculated from the \texttt{CHIMERA} retrieval using \cite{louden2017}'s method, i.e. $0.1-11.5$ at $1\sigma$ (Section \ref{freechimera}), in order for the comparison to be consistent. The atmospheric metallicities estimated from the two free-chemistry retrievals are at $<1.3 \sigma$ agreement with each other (Table \ref{tabres}, third column)  and with the metallicity retrieved in the \texttt{CHIMERA} equilibrium chemistry model. Small differences in retrieved properties for WASP-52b between \texttt{CHIMERA} and \texttt{NEMESIS} are expected, given differences in the underlying framework and approaches. However, the two methods provide statistically consistent descriptions of WASP-52b's atmosphere, which gives us confidence in the interpretation. To be conservative, we chose to adopt the larger error bars in the water abundance and metallicity obtained from the \texttt{CHIMERA} analysis in the remainder of our discussion about the nature of WASP-52b's atmosphere.

\section{Discussion}\label{discussion}

With its solar-like water abundance and sub-solar C/O ratio, WASP-52b's atmosphere follows the predictions of the core-accretion model for giant planets formed around solar-composition stars \citep{pollack1996,oberg2011,fortney2013,madhusudhan2014_2,madhusudhan2017,mordasini2016,espinoza2017}. In Figure \ref{massmet}, we add WASP-52b to the mass-metallicity plot for all known planets with measured atmospheric metallicity. The planet lies within the uncertainty regions of the previously derived relationship, for which we highlight the different (although constantly updated) trend between extrasolar planets and Solar System planets. As such, WASP-52b will benefit from future observations with the \textit{James Webb Space Telescope} (\textit{JWST}), which will allow a much more precise understanding of the trend.

\begin{figure}
\includegraphics[width=\columnwidth]{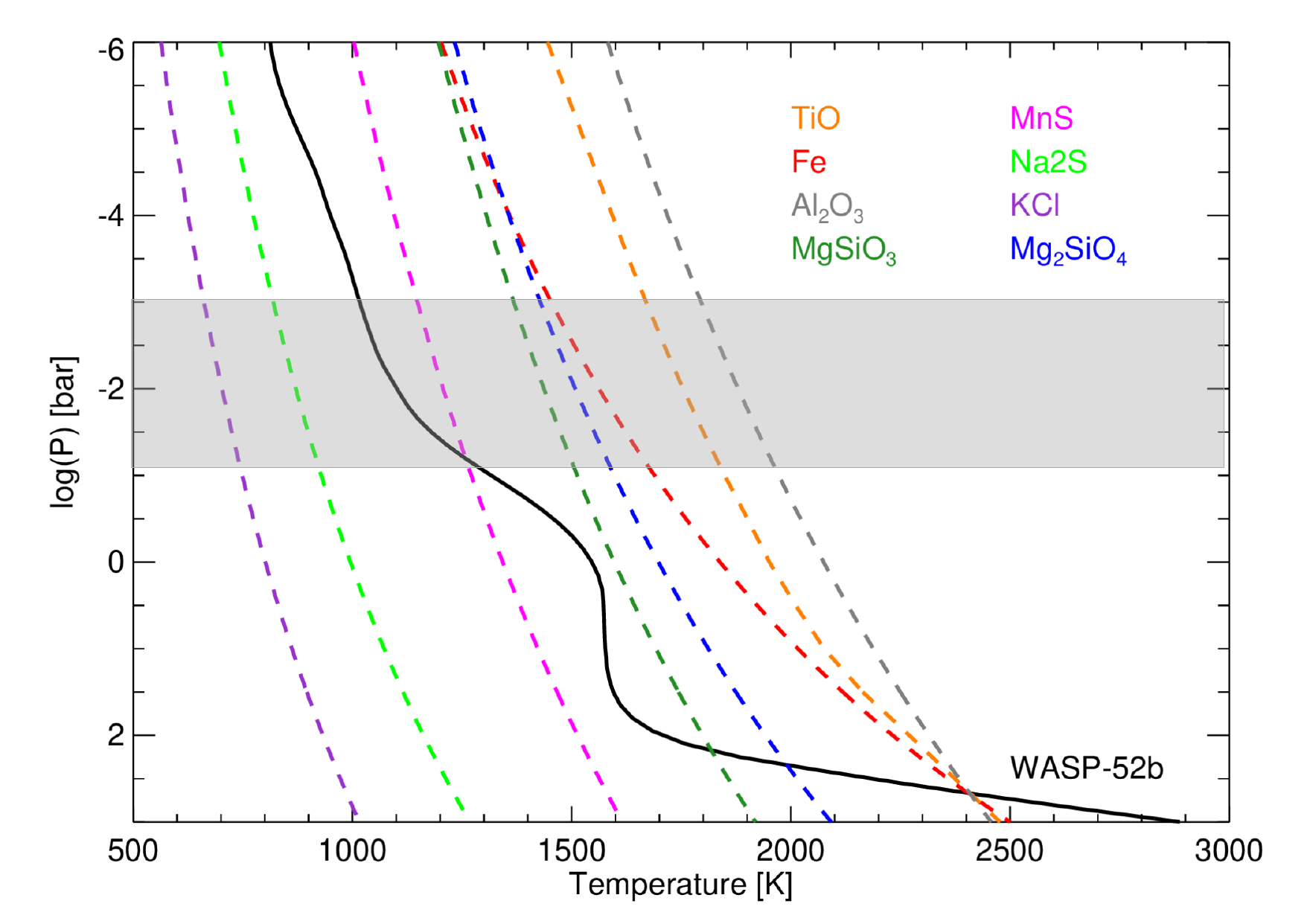}
\caption{Cloud-free, full recirculation modeled pressure-temperature profile for WASP-52b, in black. The condensation curves of several molecules in a solar composition environment are shown in dashed lines. The grey area highlights the pressure range probed by transmission spectroscopy.}
\label{PTprof}
\end{figure}

Cloud formation theory and observational trends suggest that a gas giant planet near the equilibrium temperature of WASP-52b (\teq$\sim 1300$~K) should have aerosols in the observable portion of the atmosphere \citep{barstow2017}. This is compatible with the results of our retrievals, and can explain the weak Na and K absorption features in the visible part of the spectrum. Figure \ref{PTprof} presents a one-dimensional pressure-temperature (\textit{P-T}) profile of WASP-52b, assuming a cloud-free atmosphere, efficient recirculation and an internal temperature of 1300~K (see \citealp[e.g.][]{fortney2005_tres} for further details on methodology). The intersections of the \textit{P-T} curve with condensation curves of different molecules, here calculated for a solar composition gas, inform which cloud species we would expect to form in the atmosphere of the planet and at which atmospheric depth. The deepest intersections are with silicate species (enstatite, MgSiO$_3$ and forsterite, Mg$_2$SiO$_4$), and higher in the atmosphere (around the pressures probed by transmisssion spectroscopy) are those with MnS. 

We can formulate three hypotheses for the origin of such clouds on WASP-52b. One is that their composition is MnS, and that the silicate species are confined in the deepest layers of the atmosphere. The second is that silicate species are being vertically transported into the observable portion of WASP-52b's atmosphere. Both scenarios are compatible with what we would expect given the relative cloud masses for different compositions \citep{morley2012,wakeford2018}, i.e., MnS clouds have a low relative mass compared to other species, and silicate clouds (especially enstatite) have a much larger one. This means that a relatively large amount of MnS, combined with a small contribution of silicate clouds, can explain the weak muting of the spectral features in the transmission spectrum. In the third hypothesis, we can see the spectrum as a combination of a cloud-rich planetary limb and a cloud-free one, similar to the HD 209458b atmosphere simulated by \citet{kataria2016} with a Global Circulation Model.

These three hypotheses explain our observations equally well and cannot be distinguished at the current level of observing capabilities. Future observations with \textit{JWST}, however, should be able to detect spectral features imprinted by specific aerosol species \citep{wakeford2015}.

\section{Conclusions}\label{conclusions}

We present an analysis of the visible to NIR transmission spectrum of WASP-52b, obtained with different instruments and in different epochs. As WASP-52A is a moderately active star, we relied both on ground-based photometric observations and on the modeling of the impact of stellar activity in transmission spectra. In this respect, we used analytic formulations that were previously discussed in literature and synthetic stellar spectra.

Non-occulted dark starspots during WFC3 observations can explain the relative offset between the NIR data set and the optical transmission spectrum, obtained in a different epoch. By removing such offset with an optical to IR atmospheric retrieval, and by using constraints from the ground-based photometric monitoring of the host star, we were able to estimate the average starspot temperature. When fixing the starspot temperature to values provided by literature for a K2V star, we retrieved the fraction of the stellar photosphere that is covered in starspots. In this way, we avoided simultaneously using parameters which are at least partially degenerate, and constrained each one of them with information coming from previous observations. Other than using different starspot correction approaches, we compared results obtained when applying the activity correction to the visible or to the infrared part of the transmission spectrum. The need of stitching the spectra together provided enough constraint to complement the uncertainty of the activity corrections carried out by \cite{alam2018} and \cite{bruno2018_w52} on the two separate data sets.

That being said, our results are likely affected by the systematic models used to obtain the optical and IR transmission spectra from the \textit{HST} spectroscopic transits in \cite{alam2018} and \cite{bruno2018_w52}. In the same way, we cannot exclude biases related to the uncertainty in the measure of the out-of-transit stellar flux, despite the photometric follow-up. Therefore, we could only place a lower limit on the starspot temperature or, alternatively, an upper limit on the starspot fractional coverage. This can explain why the measured starspot temperature is lower than expected for a K-type star. However, the measure of the starspot filling factor follows predictions for such a stellar type.

Our retrievals showed WASP-52b to be an ``ordinary'' hot Jupiter. With a composition compatible with solar and sub-solar C/O, this planet agrees with the mass-metallicity trend derived for the other giant planets for which this measurement is available. Our results are consistent with weak muting of the spectral features by clouds, which is compatible with a few scenarios for aerosol species. We performed retrieval exercises with two different atmospheric codes, with which we obtained results in agreement at the $\lesssim 2 \sigma$ level. 
Small differences are expected, given differences in the underlying framework and approaches, and we highlight the need of further analyses on this topic. 

Despite the broad wavelength coverage, our retrievals returned large uncertainties on the metallicity, C/O ratio and individual molecular abundances in WASP-52b's atmosphere, as well as on the contribution of clouds. However, we could exclude significant correlations between these and the starspot parameters. 

To conclude, we highlight that adding the stellar contamination to obtain a ``good fit'' of a transmission spectrum might hide spectral features that are planetary in nature, as long as they can be explained by stellar features. This implies that possible limitations of atmospheric models might be missed, and explained with the signal introduced by the star. Hence, the approach of explaining features we do not understand with stellar activity has its risks. In this respect, future observations of WASP-52b with \textit{JWST}, which have been planned by the NIRSpec Guaranteed-Time Observer (GTO) team, will help to add important constraints on the characteristics of this apparently ordinary hot Jupiter. 

\section*{ORCID iDs}
Giovanni Bruno: 0000-0002-3288-0802\\
Nikole K. Lewis: 0000-0002-8507-1304\\
Munazza K. Alam: 0000-0003-4157-832X\\
Mercedes L\'opez-Morales: 0000-0003-
3204-8183\\
Joanna K. Barstow: 0000-0003-3726-5419\\
Hannah R. Wakeford: 0000-0003-
4328-3867\\
David Sing: 0000-0001-6050-7645\\
Gregory W. Henry: 0000-0003-4155-8513\\
Vincent Bourrier: 0000-0002-9148-034X\\
Gilda E. Ballester: 0000-0002-3891-7645\\
Lars A. Buchhave: 0000-0003-1605-5666\\
Ofer Cohen: 0000-0003-3721-0215\\
Thomas M. Evans: 0000-0001-5442-1300\\
Antonio Garc\'ia Mu\~noz: 0000-0003-1756-4825\\
Panayotis Lavvas: 0000-0002-5360-3660\\
Jorge Sanz-Forcada: 0000-0002-1600-7835

\section*{Acknowledgements}

Based on observations made with the NASA/ESA Hubble Space Telescope, obtained from the data archive at the Space Telescope Science Institute. STScI is operated by the
Association of Universities for Research in Astronomy, Inc. under NASA contract NAS 5-26555. These observations are associated with program GO 14767 and GO 14260. MKA acknowledges support by the National Science Foundation through a Graduate Research Fellowship. JKB acknowledges the support of a Royal Astronomical Society Research Fellowship. VB acknowledges support by the Swiss National Science Foundation (SNSF) in the frame of the National Centre for Competence in Research ``PlanetS'', and has received funding from the European Research Council (ERC) under the European Union's Horizon 2020 research and innovation programme (project Four Aces; grant agreement No 724427). AGM acknowledges the support of the DFG priority program SPP 1992 ``Exploring the Diversity of Extrasolar Planets'' (GA 2557/1-1). The authors thank Michael R. Line and Natasha Batalha for their help with the use of CHIMERA. We thank Patrick Irwin for the use of the NEMESIS code and Ryan MacDonald for helpful suggestions on the clarity of the text. Figures \ref{Tspot_corner} and \ref{nemesis} were made with the \texttt{corner.py} package \citep{corner}.




\bibliographystyle{mnras}
\bibliography{biblio} 

\begin{thebibliography}{}
\makeatletter
\relax
\def\mn@urlcharsother{\let\do\@makeother \do\$\do\&\do\#\do\^\do\_\do\%\do\~}
\def\mn@doi{\begingroup\mn@urlcharsother \@ifnextchar [ {\mn@doi@}
  {\mn@doi@[]}}
\def\mn@doi@[#1]#2{\def\@tempa{#1}\ifx\@tempa\@empty \href
  {http://dx.doi.org/#2} {doi:#2}\else \href {http://dx.doi.org/#2} {#1}\fi
  \endgroup}
\def\mn@eprint#1#2{\mn@eprint@#1:#2::\@nil}
\def\mn@eprint@arXiv#1{\href {http://arxiv.org/abs/#1} {{\tt arXiv:#1}}}
\def\mn@eprint@dblp#1{\href {http://dblp.uni-trier.de/rec/bibtex/#1.xml}
  {dblp:#1}}
\def\mn@eprint@#1:#2:#3:#4\@nil{\def\@tempa {#1}\def\@tempb {#2}\def\@tempc
  {#3}\ifx \@tempc \@empty \let \@tempc \@tempb \let \@tempb \@tempa \fi \ifx
  \@tempb \@empty \def\@tempb {arXiv}\fi \@ifundefined
  {mn@eprint@\@tempb}{\@tempb:\@tempc}{\expandafter \expandafter \csname
  mn@eprint@\@tempb\endcsname \expandafter{\@tempc}}}

\bibitem[\protect\citeauthoryear{{Affer}, {Micela}, {Favata}  \&
  {Flaccomio}}{{Affer} et~al.}{2012}]{affer2012}
{Affer} L.,  {Micela} G.,  {Favata} F.,   {Flaccomio} E.,  2012, \mn@doi
  [\mnras] {10.1111/j.1365-2966.2012.20802.x}, \href
  {http://adsabs.harvard.edu/abs/2012MNRAS.424...11A} {424, 11}

\bibitem[\protect\citeauthoryear{{Aigrain}, {Pont}  \& {Zucker}}{{Aigrain}
  et~al.}{2012}]{aigrain2012}
{Aigrain} S.,  {Pont} F.,   {Zucker} S.,  2012, \mn@doi [\mnras]
  {10.1111/j.1365-2966.2011.19960.x}, \href
  {http://adsabs.harvard.edu/abs/2012MNRAS.419.3147A} {419, 3147}

\bibitem[\protect\citeauthoryear{{Alam} et~al.,}{{Alam}
  et~al.}{2018}]{alam2018}
{Alam} M.~K.,  et~al., 2018, \mn@doi [\aj] {10.3847/1538-3881/aaee89}, \href
  {https://ui.adsabs.harvard.edu/\#abs/2018AJ....156..298A} {156, 298}

\bibitem[\protect\citeauthoryear{{Asplund}, {Grevesse}, {Sauval}  \&
  {Scott}}{{Asplund} et~al.}{2009}]{asplund2009}
{Asplund} M.,  {Grevesse} N.,  {Sauval} A.~J.,   {Scott} P.,  2009, \mn@doi
  [Annual Review of Astronomy and Astrophysics]
  {10.1146/annurev.astro.46.060407.145222}, \href
  {https://ui.adsabs.harvard.edu/#abs/2009ARA&A..47..481A} {47, 481}

\bibitem[\protect\citeauthoryear{{Ballerini}, {Micela}, {Lanza}  \&
  {Pagano}}{{Ballerini} et~al.}{2012}]{ballerini2012}
{Ballerini} P.,  {Micela} G.,  {Lanza} A.~F.,   {Pagano} I.,  2012, \mn@doi
  [\aap] {10.1051/0004-6361/201117102}, \href
  {http://adsabs.harvard.edu/abs/2012A%26A...539A.140B} {539, A140}

\bibitem[\protect\citeauthoryear{{Barstow}, {Aigrain}, {Irwin}, {Kendrew}  \&
  {Fletcher}}{{Barstow} et~al.}{2015}]{barstow2015}
{Barstow} J.~K.,  {Aigrain} S.,  {Irwin} P.~G.~J.,  {Kendrew} S.,   {Fletcher}
  L.~N.,  2015, \mn@doi [\mnras] {10.1093/mnras/stv186}, \href
  {http://adsabs.harvard.edu/abs/2015MNRAS.448.2546B} {448, 2546}

\bibitem[\protect\citeauthoryear{{Barstow}, {Aigrain}, {Irwin}  \&
  {Sing}}{{Barstow} et~al.}{2017}]{barstow2017}
{Barstow} J.~K.,  {Aigrain} S.,  {Irwin} P.~G.~J.,   {Sing} D.~K.,  2017,
  \mn@doi [\apj] {10.3847/1538-4357/834/1/50}, \href
  {https://ui.adsabs.harvard.edu/\#abs/2017ApJ...834...50B} {834, 50}

\bibitem[\protect\citeauthoryear{{Barstow}, {Garland}, {Line}, {Rocchetto}  \&
  {Waldmann}}{{Barstow} et~al.}{2018}]{barstow2018}
{Barstow} J.,  {Garland} R.,  {Line} M.,  {Rocchetto} M.,   {Waldmann} I.,
  2018, in European Planetary Science Congress. pp EPSC2018--467

\bibitem[\protect\citeauthoryear{{Basri}, {Walkowicz}  \& {Reiners}}{{Basri}
  et~al.}{2013}]{basri2013}
{Basri} G.,  {Walkowicz} L.~M.,   {Reiners} A.,  2013, \mn@doi [\apj]
  {10.1088/0004-637X/769/1/37}, \href
  {https://ui.adsabs.harvard.edu/abs/2013ApJ...769...37B} {769, 37}

\bibitem[\protect\citeauthoryear{{Benneke} \& {Seager}}{{Benneke} \&
  {Seager}}{2012}]{benneke2012}
{Benneke} B.,  {Seager} S.,  2012, \mn@doi [\apj]
  {10.1088/0004-637X/753/2/100}, \href
  {http://adsabs.harvard.edu/abs/2012ApJ...753..100B} {753, 100}

\bibitem[\protect\citeauthoryear{{Benneke} et~al.,}{{Benneke}
  et~al.}{2019}]{benneke2019}
{Benneke} B.,  et~al., 2019, \mn@doi [Nature Astronomy]
  {10.1038/s41550-019-0800-5}, \href
  {https://ui.adsabs.harvard.edu/abs/2019NatAs...3..813B} {3, 813}

\bibitem[\protect\citeauthoryear{{Berdyugina}}{{Berdyugina}}{2005}]{berdyugina2005}
{Berdyugina} S.~V.,  2005, \mn@doi [Living Reviews in Solar Physics]
  {10.12942/lrsp-2005-8}, \href
  {http://adsabs.harvard.edu/abs/2005LRSP....2....8B} {2, 8}

\bibitem[\protect\citeauthoryear{{Borysow}}{{Borysow}}{2002}]{borysow2002}
{Borysow} A.,  2002, \mn@doi [\aap] {10.1051/0004-6361:20020555}, \href
  {https://ui.adsabs.harvard.edu/\#abs/2002A&A...390..779B} {390, 779}

\bibitem[\protect\citeauthoryear{{Borysow} \& {Frommhold}}{{Borysow} \&
  {Frommhold}}{1989}]{borysowfr1989}
{Borysow} A.,  {Frommhold} L.,  1989, \mn@doi [\apj] {10.1086/167515}, \href
  {https://ui.adsabs.harvard.edu/\#abs/1989ApJ...341..549B} {341, 549}

\bibitem[\protect\citeauthoryear{{Borysow} \& {Frommhold}}{{Borysow} \&
  {Frommhold}}{1990}]{borysowfr1990}
{Borysow} A.,  {Frommhold} L.,  1990, \mn@doi [\apj] {10.1086/185626}, \href
  {https://ui.adsabs.harvard.edu/\#abs/1990ApJ...348L..41B} {348, L41}

\bibitem[\protect\citeauthoryear{{Borysow}, {Frommhold}  \&
  {Moraldi}}{{Borysow} et~al.}{1989}]{borysowetal1989}
{Borysow} A.,  {Frommhold} L.,   {Moraldi} M.,  1989, \mn@doi [\apj]
  {10.1086/167027}, \href
  {https://ui.adsabs.harvard.edu/\#abs/1989ApJ...336..495B} {336, 495}

\bibitem[\protect\citeauthoryear{{Borysow}, {Jorgensen}  \& {Zheng}}{{Borysow}
  et~al.}{1997}]{borysow1997}
{Borysow} A.,  {Jorgensen} U.~G.,   {Zheng} C.,  1997, \aap, \href
  {https://ui.adsabs.harvard.edu/\#abs/1997A&A...324..185B} {324, 185}

\bibitem[\protect\citeauthoryear{{Bruno} et~al.,}{{Bruno}
  et~al.}{2016}]{bruno2016}
{Bruno} G.,  et~al., 2016, \mn@doi [\aap] {10.1051/0004-6361/201527699}, \href
  {http://adsabs.harvard.edu/abs/2016A%26A...595A..89B} {595, A89}

\bibitem[\protect\citeauthoryear{{Bruno} et~al.,}{{Bruno}
  et~al.}{2018}]{bruno2018_w52}
{Bruno} G.,  et~al., 2018, \mn@doi [\aj] {10.3847/1538-3881/aac6db}, \href
  {https://ui.adsabs.harvard.edu/#abs/2018AJ....156..124B} {156, 124}

\bibitem[\protect\citeauthoryear{{Buchner} et~al.,}{{Buchner}
  et~al.}{2014}]{buchner2014}
{Buchner} J.,  et~al., 2014, \mn@doi [\aap] {10.1051/0004-6361/201322971},
  \href {http://adsabs.harvard.edu/abs/2014A%26A...564A.125B} {564, A125}

\bibitem[\protect\citeauthoryear{{Chachan} et~al.,}{{Chachan}
  et~al.}{2019}]{chachan2019}
{Chachan} Y.,  et~al., 2019, \mn@doi [\aj] {10.3847/1538-3881/ab4e9a}, \href
  {https://ui.adsabs.harvard.edu/abs/2019AJ....158..244C} {158, 244}

\bibitem[\protect\citeauthoryear{{Chen}, {Pall{\'e}}, {Nortmann}, {Murgas},
  {Parviainen}  \& {Nowak}}{{Chen} et~al.}{2017}]{chen2017_w52}
{Chen} G.,  {Pall{\'e}} E.,  {Nortmann} L.,  {Murgas} F.,  {Parviainen} H.,
  {Nowak} G.,  2017, \mn@doi [\aap] {10.1051/0004-6361/201730736}, \href
  {https://ui.adsabs.harvard.edu/abs/2017A&A...600L..11C} {600, L11}

\bibitem[\protect\citeauthoryear{{Czesla}, {Huber}, {Wolter}, {Schr{\"o}ter}
  \& {Schmitt}}{{Czesla} et~al.}{2009}]{czesla2009}
{Czesla} S.,  {Huber} K.~F.,  {Wolter} U.,  {Schr{\"o}ter} S.,   {Schmitt}
  J.~H.~M.~M.,  2009, \mn@doi [\aap] {10.1051/0004-6361/200912454}, \href
  {http://adsabs.harvard.edu/abs/2009A%26A...505.1277C} {505, 1277}

\bibitem[\protect\citeauthoryear{{D{\'e}sert} et~al.,}{{D{\'e}sert}
  et~al.}{2011}]{desert2011}
{D{\'e}sert} J.-M.,  et~al., 2011, \mn@doi [\apjs]
  {10.1088/0067-0049/197/1/14}, \href
  {http://adsabs.harvard.edu/abs/2011ApJS..197...14D} {197, 14}

\bibitem[\protect\citeauthoryear{{Eaton}, {Henry}  \& {Fekel}}{{Eaton}
  et~al.}{1996}]{eaton1996}
{Eaton} J.~A.,  {Henry} G.~W.,   {Fekel} F.~C.,  1996, \mn@doi [\apj]
  {10.1086/177202}, \href
  {https://ui.adsabs.harvard.edu/abs/1996ApJ...462..888E} {462, 888}

\bibitem[\protect\citeauthoryear{{Eaton}, {Henry}  \& {Fekel}}{{Eaton}
  et~al.}{2003}]{eaton2003}
{Eaton} J.~A.,  {Henry} G.~W.,   {Fekel} F.~C.,  2003, in {Oswalt} T.~D.,  ed.,
   Vol. 288, Astrophysics and Space Science Library. p.~189,
  \mn@doi{10.1007/978-94-010-0253-0_38}

\bibitem[\protect\citeauthoryear{{Espinoza}, {Fortney}, {Miguel}, {Thorngren}
  \& {Murray-Clay}}{{Espinoza} et~al.}{2017}]{espinoza2017}
{Espinoza} N.,  {Fortney} J.~J.,  {Miguel} Y.,  {Thorngren} D.,   {Murray-Clay}
  R.,  2017, \mn@doi [\apj] {10.3847/2041-8213/aa65ca}, \href
  {https://ui.adsabs.harvard.edu/\#abs/2017ApJ...838L...9E} {838, L9}

\bibitem[\protect\citeauthoryear{{Fekel} \& {Henry}}{{Fekel} \&
  {Henry}}{2005}]{fekel2005}
{Fekel} F.~C.,  {Henry} G.~W.,  2005, \mn@doi [\aj] {10.1086/427713}, \href
  {https://ui.adsabs.harvard.edu/abs/2005AJ....129.1669F} {129, 1669}

\bibitem[\protect\citeauthoryear{{Feroz} \& {Hobson}}{{Feroz} \&
  {Hobson}}{2008}]{feroz2008}
{Feroz} F.,  {Hobson} M.~P.,  2008, \mn@doi [\mnras]
  {10.1111/j.1365-2966.2007.12353.x}, \href
  {https://ui.adsabs.harvard.edu/\#abs/2008MNRAS.384..449F} {384, 449}

\bibitem[\protect\citeauthoryear{{Feroz}, {Hobson}  \& {Bridges}}{{Feroz}
  et~al.}{2009}]{feroz2009}
{Feroz} F.,  {Hobson} M.~P.,   {Bridges} M.,  2009, \mn@doi [\mnras]
  {10.1111/j.1365-2966.2009.14548.x}, \href
  {http://adsabs.harvard.edu/abs/2009MNRAS.398.1601F} {398, 1601}

\bibitem[\protect\citeauthoryear{Foreman-Mackey}{Foreman-Mackey}{2016}]{corner}
Foreman-Mackey D.,  2016, \mn@doi [The Journal of Open Source Software]
  {10.21105/joss.00024}, 24

\bibitem[\protect\citeauthoryear{{Fortney}, {Marley}, {Lodders}, {Saumon}  \&
  {Freedman}}{{Fortney} et~al.}{2005}]{fortney2005_tres}
{Fortney} J.~J.,  {Marley} M.~S.,  {Lodders} K.,  {Saumon} D.,   {Freedman} R.,
   2005, \mn@doi [\apjl] {10.1086/431952}, \href
  {http://adsabs.harvard.edu/abs/2005ApJ...627L..69F} {627, L69}

\bibitem[\protect\citeauthoryear{{Fortney}, {Mordasini}, {Nettelmann},
  {Kempton}, {Greene}  \& {Zahnle}}{{Fortney} et~al.}{2013}]{fortney2013}
{Fortney} J.~J.,  {Mordasini} C.,  {Nettelmann} N.,  {Kempton} E. M.~R.,
  {Greene} T.~P.,   {Zahnle} K.,  2013, \mn@doi [\apj]
  {10.1088/0004-637X/775/1/80}, \href
  {https://ui.adsabs.harvard.edu/\#abs/2013ApJ...775...80F} {775, 80}

\bibitem[\protect\citeauthoryear{{Fu}, {Deming}, {Knutson}, {Madhusudhan},
  {Mandell}  \& {Fraine}}{{Fu} et~al.}{2017}]{fu2017}
{Fu} G.,  {Deming} D.,  {Knutson} H.,  {Madhusudhan} N.,  {Mandell} A.,
  {Fraine} J.,  2017, \mn@doi [\apjl] {10.3847/2041-8213/aa8e40}, \href
  {https://ui.adsabs.harvard.edu/abs/2017ApJ...847L..22F} {847, L22}

\bibitem[\protect\citeauthoryear{{H{\'e}brard} et~al.,}{{H{\'e}brard}
  et~al.}{2013}]{hebrard2013_w52}
{H{\'e}brard} G.,  et~al., 2013, \mn@doi [\aap] {10.1051/0004-6361/201220363},
  \href {http://adsabs.harvard.edu/abs/2013A%26A...549A.134H} {549, A134}

\bibitem[\protect\citeauthoryear{{Heiter} et~al.,}{{Heiter}
  et~al.}{2008}]{heiter2008}
{Heiter} U.,  et~al., 2008, in Journal of Physics Conference Series. p. 012011,
  \mn@doi{10.1088/1742-6596/130/1/012011}

\bibitem[\protect\citeauthoryear{{Heng}}{{Heng}}{2017}]{heng2017book}
{Heng} K.,  2017, {Exoplanetary Atmospheres: Theoretical Concepts and
  Foundations}

\bibitem[\protect\citeauthoryear{{Henry}}{{Henry}}{1999}]{henry1999}
{Henry} G.~W.,  1999, \mn@doi [\pasp] {10.1086/316388}, \href
  {https://ui.adsabs.harvard.edu/abs/1999PASP..111..845H} {111, 845}

\bibitem[\protect\citeauthoryear{{Huitson} et~al.,}{{Huitson}
  et~al.}{2013}]{huitson2013}
{Huitson} C.~M.,  et~al., 2013, \mn@doi [\mnras] {10.1093/mnras/stt1243}, \href
  {http://adsabs.harvard.edu/abs/2013MNRAS.434.3252H} {434, 3252}

\bibitem[\protect\citeauthoryear{{Husser}, {Wende-von Berg}, {Dreizler},
  {Homeier}, {Reiners}, {Barman}  \& {Hauschildt}}{{Husser}
  et~al.}{2013}]{husser2013}
{Husser} T.-O.,  {Wende-von Berg} S.,  {Dreizler} S.,  {Homeier} D.,  {Reiners}
  A.,  {Barman} T.,   {Hauschildt} P.~H.,  2013, \mn@doi [\aap]
  {10.1051/0004-6361/201219058}, \href
  {http://adsabs.harvard.edu/abs/2013A%26A...553A...6H} {553, A6}

\bibitem[\protect\citeauthoryear{Jones, Oliphant, Peterson  et~al.}{Jones
  et~al.}{2001}]{scipy}
Jones E.,  Oliphant T.,  Peterson P.,   et~al., 2001, {SciPy}: Open source
  scientific tools for {Python}, \url {http://www.scipy.org/}

\bibitem[\protect\citeauthoryear{{Kajatkari}, {Hackman}, {Jetsu}, {Lehtinen}
  \& {Henry}}{{Kajatkari} et~al.}{2014}]{kajatkari2014}
{Kajatkari} P.,  {Hackman} T.,  {Jetsu} L.,  {Lehtinen} J.,   {Henry} G.~W.,
  2014, \mn@doi [\aap] {10.1051/0004-6361/201321291}, \href
  {https://ui.adsabs.harvard.edu/abs/2014A&A...562A.107K} {562, A107}

\bibitem[\protect\citeauthoryear{{Kataria}, {Sing}, {Lewis}, {Visscher},
  {Showman}, {Fortney}  \& {Marley}}{{Kataria} et~al.}{2016}]{kataria2016}
{Kataria} T.,  {Sing} D.~K.,  {Lewis} N.~K.,  {Visscher} C.,  {Showman} A.~P.,
  {Fortney} J.~J.,   {Marley} M.~S.,  2016, \mn@doi [\apj]
  {10.3847/0004-637X/821/1/9}, \href
  {https://ui.adsabs.harvard.edu/\#abs/2016ApJ...821....9K} {821, 9}

\bibitem[\protect\citeauthoryear{{Kilpatrick} et~al.,}{{Kilpatrick}
  et~al.}{2018}]{kilpatrick2018}
{Kilpatrick} B.~M.,  et~al., 2018, \mn@doi [\aj] {10.3847/1538-3881/aacea7},
  \href {https://ui.adsabs.harvard.edu/\#abs/2018AJ....156..103K} {156, 103}

\bibitem[\protect\citeauthoryear{{Kirk}, {Wheatley}, {Louden}, {Littlefair},
  {Copperwheat}, {Armstrong}, {Marsh}  \& {Dhillon}}{{Kirk}
  et~al.}{2016}]{kirk2016}
{Kirk} J.,  {Wheatley} P.~J.,  {Louden} T.,  {Littlefair} S.~P.,  {Copperwheat}
  C.~M.,  {Armstrong} D.~J.,  {Marsh} T.~R.,   {Dhillon} V.~S.,  2016, \mn@doi
  [\mnras] {10.1093/mnras/stw2205}, \href
  {http://adsabs.harvard.edu/abs/2016MNRAS.463.2922K} {463, 2922}

\bibitem[\protect\citeauthoryear{{Kreidberg} et~al.,}{{Kreidberg}
  et~al.}{2014}]{kreidberg2014}
{Kreidberg} L.,  et~al., 2014, \mn@doi [\apjl] {10.1088/2041-8205/793/2/L27},
  \href {http://adsabs.harvard.edu/abs/2014ApJ...793L..27K} {793, L27}

\bibitem[\protect\citeauthoryear{{Krissansen-Totton}, {Garland}, {Irwin}  \&
  {Catling}}{{Krissansen-Totton} et~al.}{2018}]{krissansentotton2018}
{Krissansen-Totton} J.,  {Garland} R.,  {Irwin} P.,   {Catling} D.~C.,  2018,
  \mn@doi [\aj] {10.3847/1538-3881/aad564}, \href
  {https://ui.adsabs.harvard.edu/\#abs/2018AJ....156..114K} {156, 114}

\bibitem[\protect\citeauthoryear{{Lacis} \& {Oinas}}{{Lacis} \&
  {Oinas}}{1991}]{lacis1991}
{Lacis} A.~A.,  {Oinas} V.,  1991, \mn@doi [Journal of Geophysical Research]
  {10.1029/90JD01945}, \href
  {https://ui.adsabs.harvard.edu/\#abs/1991JGR....96.9027L} {96, 9027}

\bibitem[\protect\citeauthoryear{{Lecavelier Des Etangs}, {Pont},
  {Vidal-Madjar}  \& {Sing}}{{Lecavelier Des Etangs}
  et~al.}{2008}]{lecavalierdesetangs2008}
{Lecavelier Des Etangs} A.,  {Pont} F.,  {Vidal-Madjar} A.,   {Sing} D.,  2008,
  \mn@doi [\aap] {10.1051/0004-6361:200809388}, \href
  {http://adsabs.harvard.edu/abs/2008A%26A...481L..83L} {481, L83}

\bibitem[\protect\citeauthoryear{{Lehtinen}, {Jetsu}, {Hackman}, {Kajatkari}
  \& {Henry}}{{Lehtinen} et~al.}{2012}]{lehtinen2012}
{Lehtinen} J.,  {Jetsu} L.,  {Hackman} T.,  {Kajatkari} P.,   {Henry} G.~W.,
  2012, \mn@doi [\aap] {10.1051/0004-6361/201219185}, \href
  {https://ui.adsabs.harvard.edu/abs/2012A&A...542A..38L} {542, A38}

\bibitem[\protect\citeauthoryear{{Line} et~al.,}{{Line}
  et~al.}{2013}]{line2013}
{Line} M.~R.,  et~al., 2013, \mn@doi [\apj] {10.1088/0004-637X/775/2/137},
  \href {http://adsabs.harvard.edu/abs/2013ApJ...775..137L} {775, 137}

\bibitem[\protect\citeauthoryear{{Louden}, {Wheatley}, {Irwin}, {Kirk}  \&
  {Skillen}}{{Louden} et~al.}{2017}]{louden2017}
{Louden} T.,  {Wheatley} P.~J.,  {Irwin} P.~G.~J.,  {Kirk} J.,   {Skillen} I.,
  2017, \mn@doi [\mnras] {10.1093/mnras/stx984}, \href
  {http://adsabs.harvard.edu/abs/2017MNRAS.470..742L} {470, 742}

\bibitem[\protect\citeauthoryear{{Madhusudhan}}{{Madhusudhan}}{2018}]{madhusudhan2018}
{Madhusudhan} N.,  2018, {Atmospheric Retrieval of Exoplanets}.
p.~104, \mn@doi{10.1007/978-3-319-55333-7_104}

\bibitem[\protect\citeauthoryear{{Madhusudhan}, {Amin}  \&
  {Kennedy}}{{Madhusudhan} et~al.}{2014}]{madhusudhan2014_2}
{Madhusudhan} N.,  {Amin} M.~A.,   {Kennedy} G.~M.,  2014, \mn@doi [\apjl]
  {10.1088/2041-8205/794/1/L12}, \href
  {http://adsabs.harvard.edu/abs/2014ApJ...794L..12M} {794, L12}

\bibitem[\protect\citeauthoryear{{Madhusudhan}, {Bitsch}, {Johansen}  \&
  {Eriksson}}{{Madhusudhan} et~al.}{2017}]{madhusudhan2017}
{Madhusudhan} N.,  {Bitsch} B.,  {Johansen} A.,   {Eriksson} L.,  2017, \mn@doi
  [\mnras] {10.1093/mnras/stx1139}, \href
  {https://ui.adsabs.harvard.edu/\#abs/2017MNRAS.469.4102M} {469, 4102}

\bibitem[\protect\citeauthoryear{{Mancini} et~al.,}{{Mancini}
  et~al.}{2017}]{mancini2017}
{Mancini} L.,  et~al., 2017, \mn@doi [\mnras] {10.1093/mnras/stw1987}, \href
  {http://adsabs.harvard.edu/abs/2017MNRAS.465..843M} {465, 843}

\bibitem[\protect\citeauthoryear{{May}, {Zhao}, {Haidar}, {Rauscher}  \&
  {Monnier}}{{May} et~al.}{2018}]{may2018}
{May} E.~M.,  {Zhao} M.,  {Haidar} M.,  {Rauscher} E.,   {Monnier} J.~D.,
  2018, \mn@doi [\aj] {10.3847/1538-3881/aad4a8}, \href
  {https://ui.adsabs.harvard.edu/abs/2018AJ....156..122M} {156, 122}

\bibitem[\protect\citeauthoryear{{McCullough}, {Crouzet}, {Deming}  \&
  {Madhusudhan}}{{McCullough} et~al.}{2014}]{mccullough2014}
{McCullough} P.~R.,  {Crouzet} N.,  {Deming} D.,   {Madhusudhan} N.,  2014,
  \mn@doi [\apj] {10.1088/0004-637X/791/1/55}, \href
  {http://adsabs.harvard.edu/abs/2014ApJ...791...55M} {791, 55}

\bibitem[\protect\citeauthoryear{{Montalto}, {Bou{\'e}}, {Oshagh}, {Boisse},
  {Bruno}  \& {Santos}}{{Montalto} et~al.}{2014}]{montalto2014}
{Montalto} M.,  {Bou{\'e}} G.,  {Oshagh} M.,  {Boisse} I.,  {Bruno} G.,
  {Santos} N.~C.,  2014, \mn@doi [\mnras] {10.1093/mnras/stu1530}, \href
  {http://adsabs.harvard.edu/abs/2014MNRAS.444.1721M} {444, 1721}

\bibitem[\protect\citeauthoryear{{Mordasini}, {van Boekel}, {Molli{\`e}re},
  {Henning}  \& {Benneke}}{{Mordasini} et~al.}{2016}]{mordasini2016}
{Mordasini} C.,  {van Boekel} R.,  {Molli{\`e}re} P.,  {Henning} T.,
  {Benneke} B.,  2016, \mn@doi [\apj] {10.3847/0004-637X/832/1/41}, \href
  {http://adsabs.harvard.edu/abs/2016ApJ...832...41M} {832, 41}

\bibitem[\protect\citeauthoryear{{Morley}, {Fortney}, {Marley}, {Visscher},
  {Saumon}  \& {Leggett}}{{Morley} et~al.}{2012}]{morley2012}
{Morley} C.~V.,  {Fortney} J.~J.,  {Marley} M.~S.,  {Visscher} C.,  {Saumon}
  D.,   {Leggett} S.~K.,  2012, \mn@doi [\apj] {10.1088/0004-637X/756/2/172},
  \href {http://adsabs.harvard.edu/abs/2012ApJ...756..172M} {756, 172}

\bibitem[\protect\citeauthoryear{{Morley}, {Knutson}, {Line}, {Fortney},
  {Thorngren}, {Marley}, {Teal}  \& {Lupu}}{{Morley} et~al.}{2017}]{morley2017}
{Morley} C.~V.,  {Knutson} H.,  {Line} M.,  {Fortney} J.~J.,  {Thorngren} D.,
  {Marley} M.~S.,  {Teal} D.,   {Lupu} R.,  2017, \mn@doi [\aj]
  {10.3847/1538-3881/153/2/86}, \href
  {http://adsabs.harvard.edu/abs/2017AJ....153...86M} {153, 86}

\bibitem[\protect\citeauthoryear{{Moses} et~al.,}{{Moses}
  et~al.}{2011}]{moses2011}
{Moses} J.~I.,  et~al., 2011, \mn@doi [\apj] {10.1088/0004-637X/737/1/15},
  \href {https://ui.adsabs.harvard.edu/abs/2011ApJ...737...15M} {737, 15}

\bibitem[\protect\citeauthoryear{{Norris}, {Beeck}, {Unruh}, {Solanki},
  {Krivova}  \& {Yeo}}{{Norris} et~al.}{2017}]{norris2017}
{Norris} C.~M.,  {Beeck} B.,  {Unruh} Y.~C.,  {Solanki} S.~K.,  {Krivova}
  N.~A.,   {Yeo} K.~L.,  2017, \mn@doi [\aap] {10.1051/0004-6361/201629879},
  \href {http://adsabs.harvard.edu/abs/2017A%26A...605A..45N} {605, A45}

\bibitem[\protect\citeauthoryear{{{\"O}berg}, {Murray-Clay}  \&
  {Bergin}}{{{\"O}berg} et~al.}{2011}]{oberg2011}
{{\"O}berg} K.~I.,  {Murray-Clay} R.,   {Bergin} E.~A.,  2011, \mn@doi [\apj]
  {10.1088/2041-8205/743/1/L16}, \href
  {https://ui.adsabs.harvard.edu/\#abs/2011ApJ...743L..16O} {743, L16}

\bibitem[\protect\citeauthoryear{{Pinhas}, {Rackham}, {Madhusudhan}  \&
  {Apai}}{{Pinhas} et~al.}{2018}]{pinhas2018}
{Pinhas} A.,  {Rackham} B.~V.,  {Madhusudhan} N.,   {Apai} D.,  2018, \mn@doi
  [\mnras] {10.1093/mnras/sty2209}, \href
  {https://ui.adsabs.harvard.edu/abs/2018MNRAS.480.5314P} {480, 5314}

\bibitem[\protect\citeauthoryear{{Pollack}, {Hubickyj}, {Bodenheimer},
  {Lissauer}, {Podolak}  \& {Greenzweig}}{{Pollack} et~al.}{1996}]{pollack1996}
{Pollack} J.~B.,  {Hubickyj} O.,  {Bodenheimer} P.,  {Lissauer} J.~J.,
  {Podolak} M.,   {Greenzweig} Y.,  1996, \mn@doi [\icarus]
  {10.1006/icar.1996.0190}, \href
  {http://adsabs.harvard.edu/abs/1996Icar..124...62P} {124, 62}

\bibitem[\protect\citeauthoryear{{Pont} et~al.,}{{Pont}
  et~al.}{2007}]{pont2007}
{Pont} F.,  et~al., 2007, \mn@doi [\aap] {10.1051/0004-6361:20078269}, \href
  {http://adsabs.harvard.edu/abs/2007A%26A...476.1347P} {476, 1347}

\bibitem[\protect\citeauthoryear{{Prinn} \& {Barshay}}{{Prinn} \&
  {Barshay}}{1977}]{prinn1977}
{Prinn} R.~G.,  {Barshay} S.~S.,  1977, \mn@doi [Science]
  {10.1126/science.198.4321.1031}, \href
  {https://ui.adsabs.harvard.edu/abs/1977Sci...198.1031P} {198, 1031}

\bibitem[\protect\citeauthoryear{{Rackham} et~al.,}{{Rackham}
  et~al.}{2017}]{rackham2017}
{Rackham} B.,  et~al., 2017, \mn@doi [\apj] {10.3847/1538-4357/aa4f6c}, \href
  {http://adsabs.harvard.edu/abs/2017ApJ...834..151R} {834, 151}

\bibitem[\protect\citeauthoryear{{Rackham}, {Apai}  \& {Giampapa}}{{Rackham}
  et~al.}{2018}]{rackham2018}
{Rackham} B.~V.,  {Apai} D.,   {Giampapa} M.~S.,  2018, \mn@doi [\apj]
  {10.3847/1538-4357/aaa08c}, \href
  {https://ui.adsabs.harvard.edu/#abs/2018ApJ...853..122R} {853, 122}

\bibitem[\protect\citeauthoryear{{Rackham}, {Apai}  \& {Giampapa}}{{Rackham}
  et~al.}{2019}]{rackham2019}
{Rackham} B.~V.,  {Apai} D.,   {Giampapa} M.~S.,  2019, \mn@doi [\aj]
  {10.3847/1538-3881/aaf892}, \href
  {https://ui.adsabs.harvard.edu/\#abs/2019AJ....157...96R} {157, 96}

\bibitem[\protect\citeauthoryear{{Silva-Valio} \& {Lanza}}{{Silva-Valio} \&
  {Lanza}}{2011}]{silva-valio2011}
{Silva-Valio} A.,  {Lanza} A.~F.,  2011, \mn@doi [\aap]
  {10.1051/0004-6361/201015382}, \href
  {http://adsabs.harvard.edu/abs/2011A%26A...529A..36S} {529, A36}

\bibitem[\protect\citeauthoryear{{Sing} et~al.,}{{Sing}
  et~al.}{2011}]{sing2011}
{Sing} D.~K.,  et~al., 2011, \mn@doi [\mnras]
  {10.1111/j.1365-2966.2011.19142.x}, \href
  {http://adsabs.harvard.edu/abs/2011MNRAS.416.1443S} {416, 1443}

\bibitem[\protect\citeauthoryear{{Sing} et~al.,}{{Sing}
  et~al.}{2015}]{sing2015}
{Sing} D.~K.,  et~al., 2015, \mn@doi [\mnras] {10.1093/mnras/stu2279}, \href
  {http://adsabs.harvard.edu/abs/2015MNRAS.446.2428S} {446, 2428}

\bibitem[\protect\citeauthoryear{{Sing} et~al.,}{{Sing}
  et~al.}{2016}]{sing2016}
{Sing} D.~K.,  et~al., 2016, \mn@doi [\nat] {10.1038/nature16068}, \href
  {http://adsabs.harvard.edu/abs/2016Natur.529...59S} {529, 59}

\bibitem[\protect\citeauthoryear{{Stevenson}}{{Stevenson}}{2016}]{stevenson2016}
{Stevenson} K.~B.,  2016, \mn@doi [\apjl] {10.3847/2041-8205/817/2/L16}, \href
  {http://adsabs.harvard.edu/abs/2016ApJ...817L..16S} {817, L16}

\bibitem[\protect\citeauthoryear{{Thorngren}, {Fortney}, {Murray-Clay}  \&
  {Lopez}}{{Thorngren} et~al.}{2016}]{thorngren2016}
{Thorngren} D.~P.,  {Fortney} J.~J.,  {Murray-Clay} R.~A.,   {Lopez} E.~D.,
  2016, \mn@doi [\apj] {10.3847/0004-637X/831/1/64}, \href
  {http://adsabs.harvard.edu/abs/2016ApJ...831...64T} {831, 64}

\bibitem[\protect\citeauthoryear{{Wakeford} \& {Sing}}{{Wakeford} \&
  {Sing}}{2015}]{wakeford2015}
{Wakeford} H.~R.,  {Sing} D.~K.,  2015, \mn@doi [\aap]
  {10.1051/0004-6361/201424207}, \href
  {http://adsabs.harvard.edu/abs/2015A%26A...573A.122W} {573, A122}

\bibitem[\protect\citeauthoryear{{Wakeford} et~al.,}{{Wakeford}
  et~al.}{2017}]{wakeford2017}
{Wakeford} H.~R.,  et~al., 2017, \mn@doi [Science] {10.1126/science.aah4668},
  \href {https://ui.adsabs.harvard.edu/\#abs/2017Sci...356..628W} {356, 628}

\bibitem[\protect\citeauthoryear{{Wakeford} et~al.,}{{Wakeford}
  et~al.}{2018}]{wakeford2018}
{Wakeford} H.~R.,  et~al., 2018, \mn@doi [\aj] {10.3847/1538-3881/aa9e4e},
  \href {https://ui.adsabs.harvard.edu/abs/2018AJ....155...29W} {155, 29}

\bibitem[\protect\citeauthoryear{{Wakeford} et~al.,}{{Wakeford}
  et~al.}{2019}]{wakeford2019}
{Wakeford} H.~R.,  et~al., 2019, \mn@doi [\aj] {10.3847/1538-3881/aaf04d},
  \href {https://ui.adsabs.harvard.edu/abs/2019AJ....157...11W} {157, 11}

\makeatother
\end{thebibliography}




\appendix

\section{WASP-52A rotation period}\label{rphot}
\begin{figure*}
\includegraphics[width=0.8\textwidth]{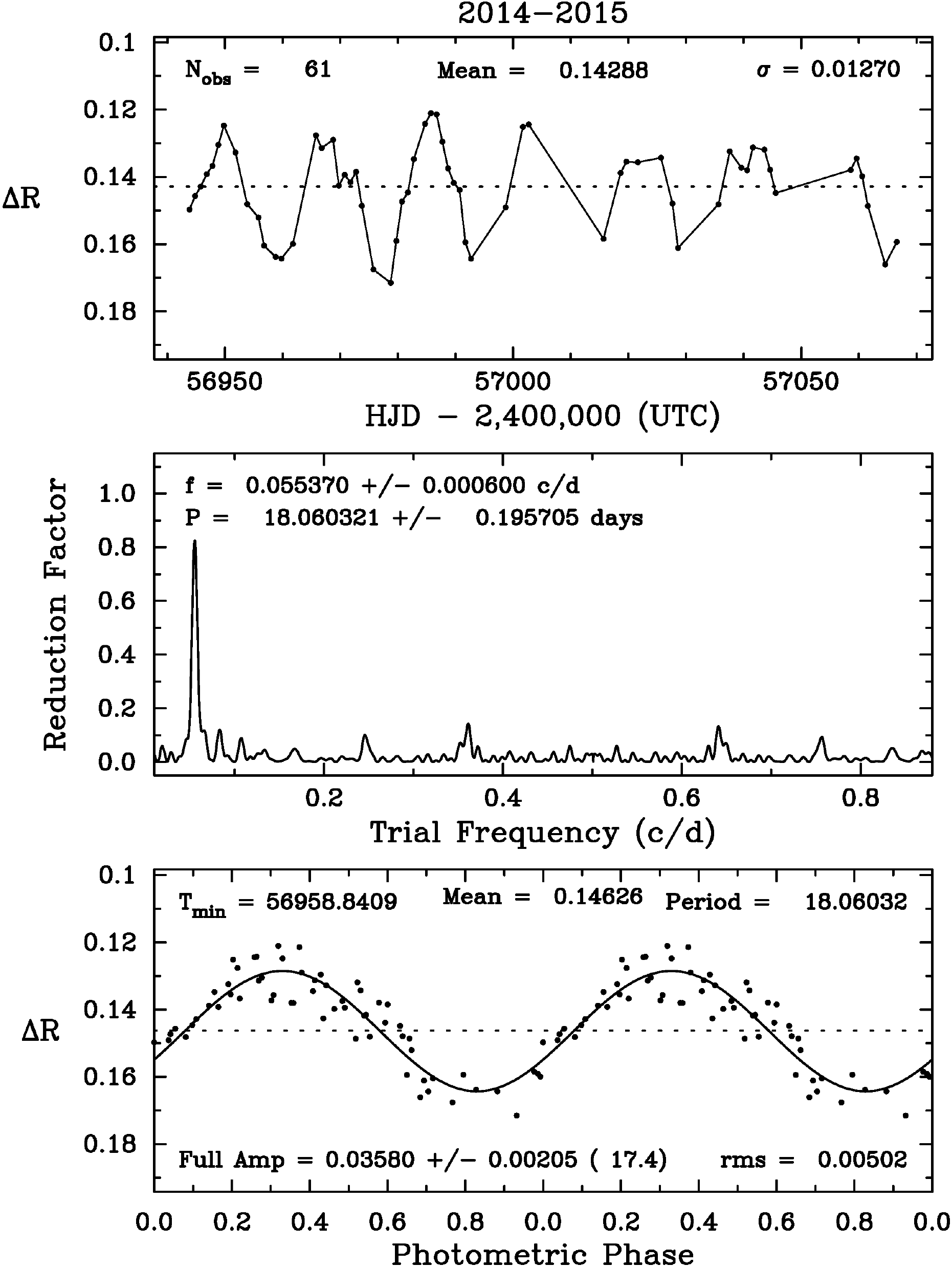}
\caption{AIT observations of WASP-52 during the 2014-2015 season. \textit{Upper panel:} Light curve in the $R$ band as a function of time; \textit{Middle panel:} periodogram showing the stellar rotation signal in cycles/day; \textit{Lower panel:} phase-folded light curve. This represents our most reliable determination of WASP-52's rotation period.}
\label{seas1}
\end{figure*}

We observed WASP-52 with the Celestron 14-inch Automated Imaging Telescope (AIT), located at Fairborn Observatory in Arizona \citep{henry1999,eaton2003}, during the 2014-15, 2015-16, 2016-17, 2017-18, and 2018-19 observing seasons. Our photometric measurements and period analysis, carried out as described in \cite{sing2015}, yielded periods between 16.7 and 18.5 days in the first four observing seasons. These values agree well with $P_\mathrm{rot} = 16.4$ days reported by \cite{hebrard2013_w52} and 17.8 days measured by \cite{louden2017}. The strongest and most coherent variability occurred in the 2014-2015 observing season (see Figure \ref{seas1}). Our measured photometric period was $18.06\pm0.20$ days, and the peak-to-peak amplitude was nearly 0.04 mag. We take this to be our best measurement of the star's rotation period.

\begin{figure*}
\includegraphics[width=0.8\textwidth]{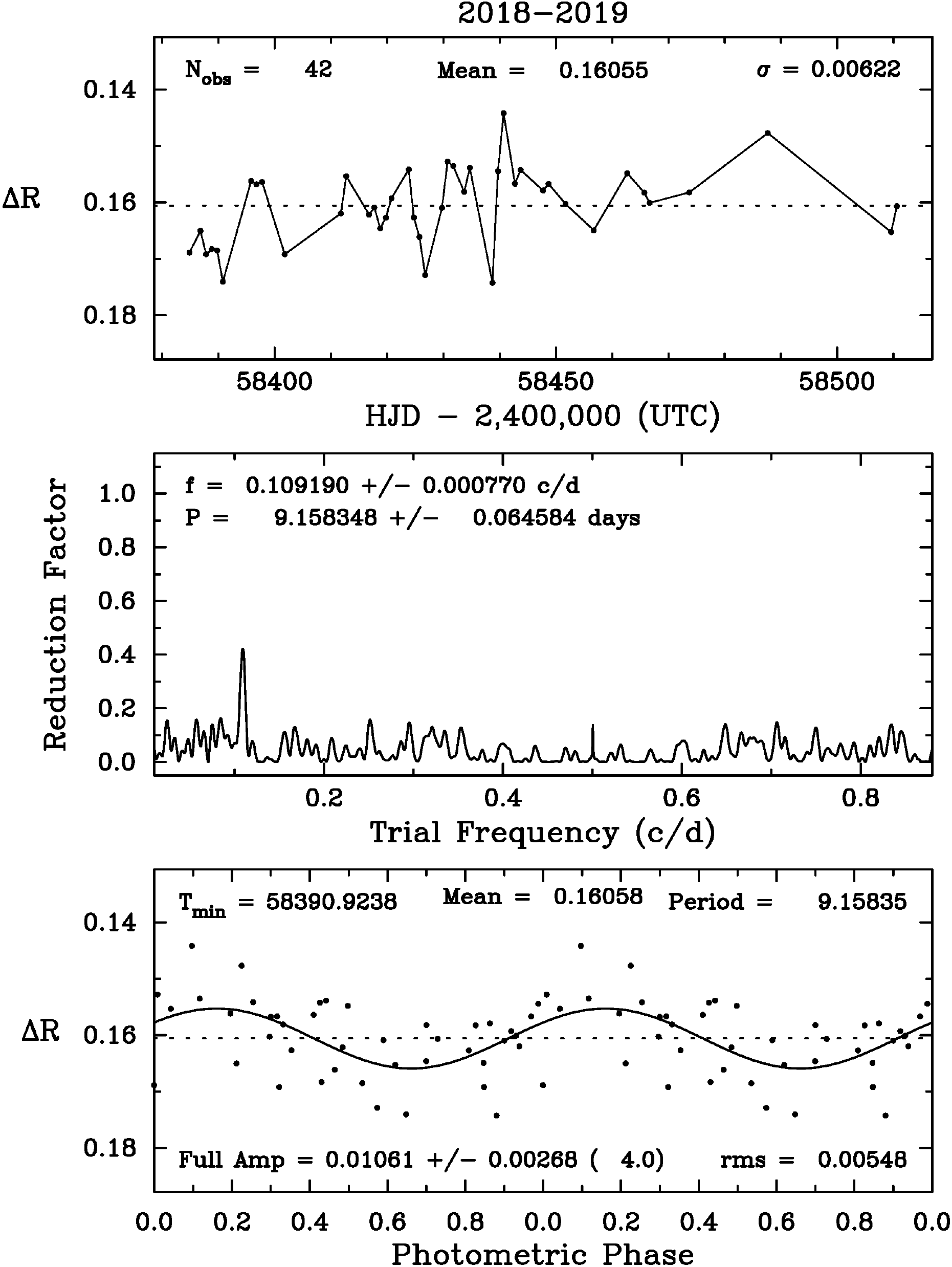}
\caption{Same as Figure \ref{seas1}, but for the 2018-2019 season. The photometric period is one half that observed in the 2014-15 observing season, implying the spot distribution on WASP-52 has evolved to place comparable spot activity on opposite hemispheres of the star.  The 9 day photometric period confirms that the 18 day period in 2014-15 is the correct stellar rotation period.}
\label{seas5}
\end{figure*}

In the fifth observing season (2018-2019), we found a period of only $9.16\pm0.06$ days and an amplitude of only 0.01 mag (see Figure \ref{seas5}). This suggests that the spot distribution changed from one hemisphere dominating to comparable spot activity on opposite hemispheres \citep[e.g.][]{eaton1996,fekel2005,lehtinen2012,kajatkari2014}. Detection of the 9-day period confirms that 18 days is the star's correct rotation period.

\section{Retrieval plots}

\begin{figure*}
\includegraphics[width=\textwidth]{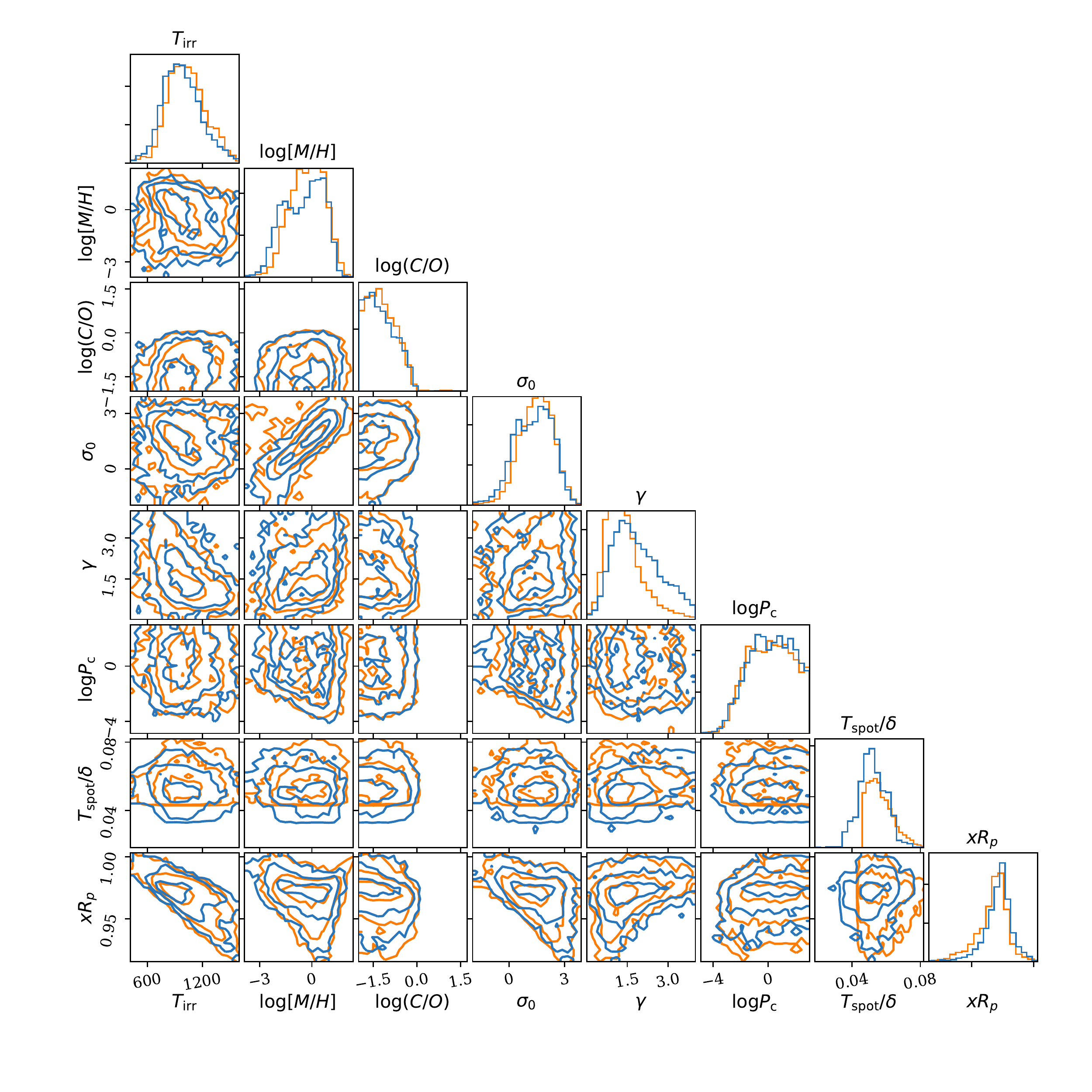}
\caption{Marginalized posterior distributions obtained with \texttt{CHIMERA}. Shown are the $1$, $2$ and $3\sigma$ credible intervals for the retrievals with a dark spot acting on the WFC3 data set. The retrieval using the starspot filling factor (blue, $\delta$) and the starspot temperature (orange, $T_\mathrm{spot}$) are compared. The starspot temperature was scaled to the $\delta$ parameter prior with the relationship $T_\mathrm{spot} \rightarrow T_\mathrm{spot}/(10000 \, \mathrm{K}) - 0.23$ (so that $2300 \, \mathrm{K}$ corresponds to 0 and $3000 \, \mathrm{K}$ corresponds to 0.07)} and is indicated as $T_\mathrm{spot}$ in the same panels as $\delta$.
\label{Tspot_corner}
\end{figure*}

\begin{figure*}
\includegraphics[width=\textwidth]{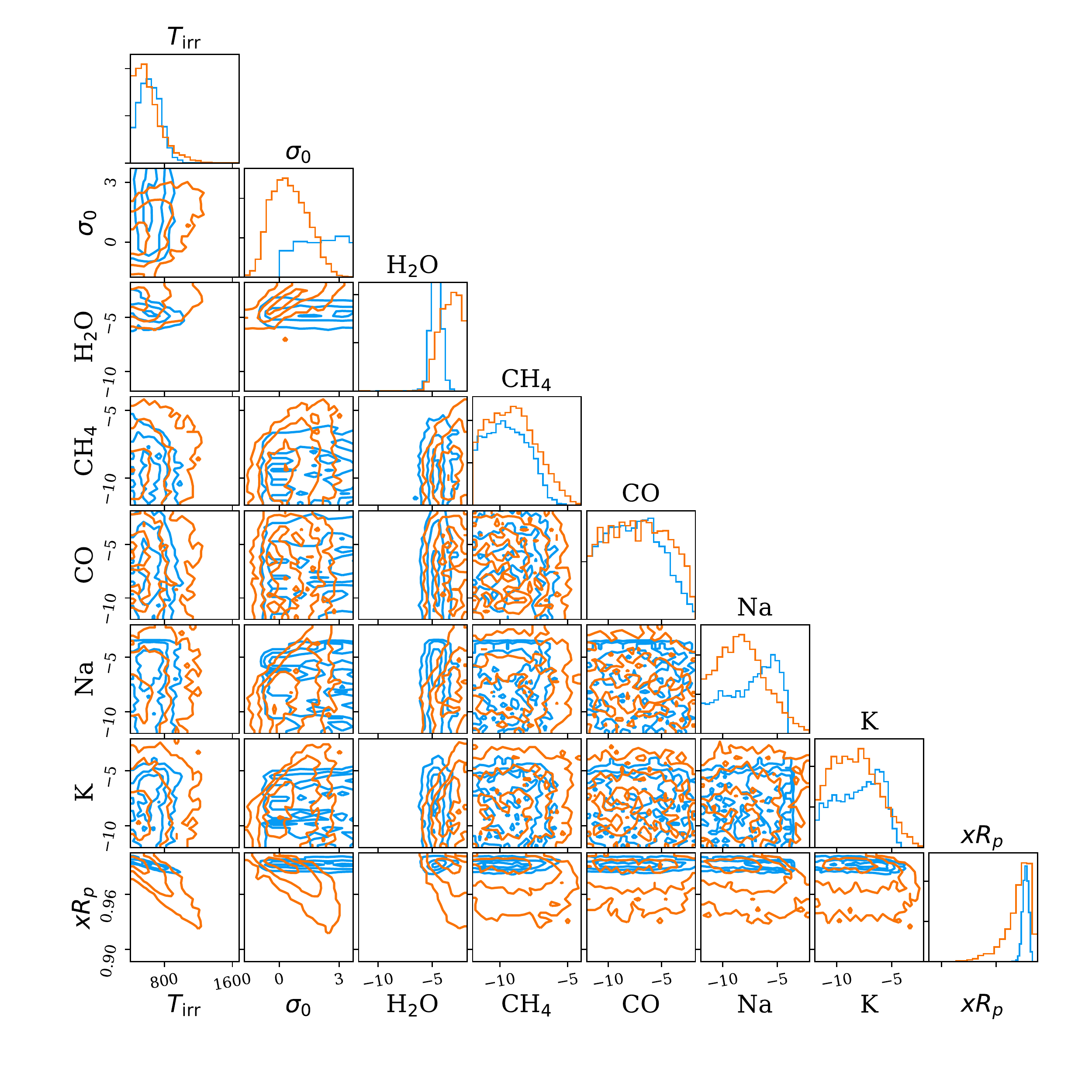}
\caption{Marginalized posterior distributions obtained with \texttt{CHIMERA} (orange) and \texttt{NEMESIS} (blue) compared. The $1$, $2$ and $3\sigma$ credible intervals are only shown for parameters with similar meaning, i.e. temperature, scattering index, molecular abundances and radius scaling factor. Because of the difference among retrieved models, neither scattering cross section or starspot temperature (\texttt{CHIMERA} only), nor cloud parameters (treated differently between \texttt{CHIMERA} and \texttt{NEMESIS}) are shown.}
\label{nemesis}
\end{figure*}

\begin{figure*}
\includegraphics[width=\textwidth]{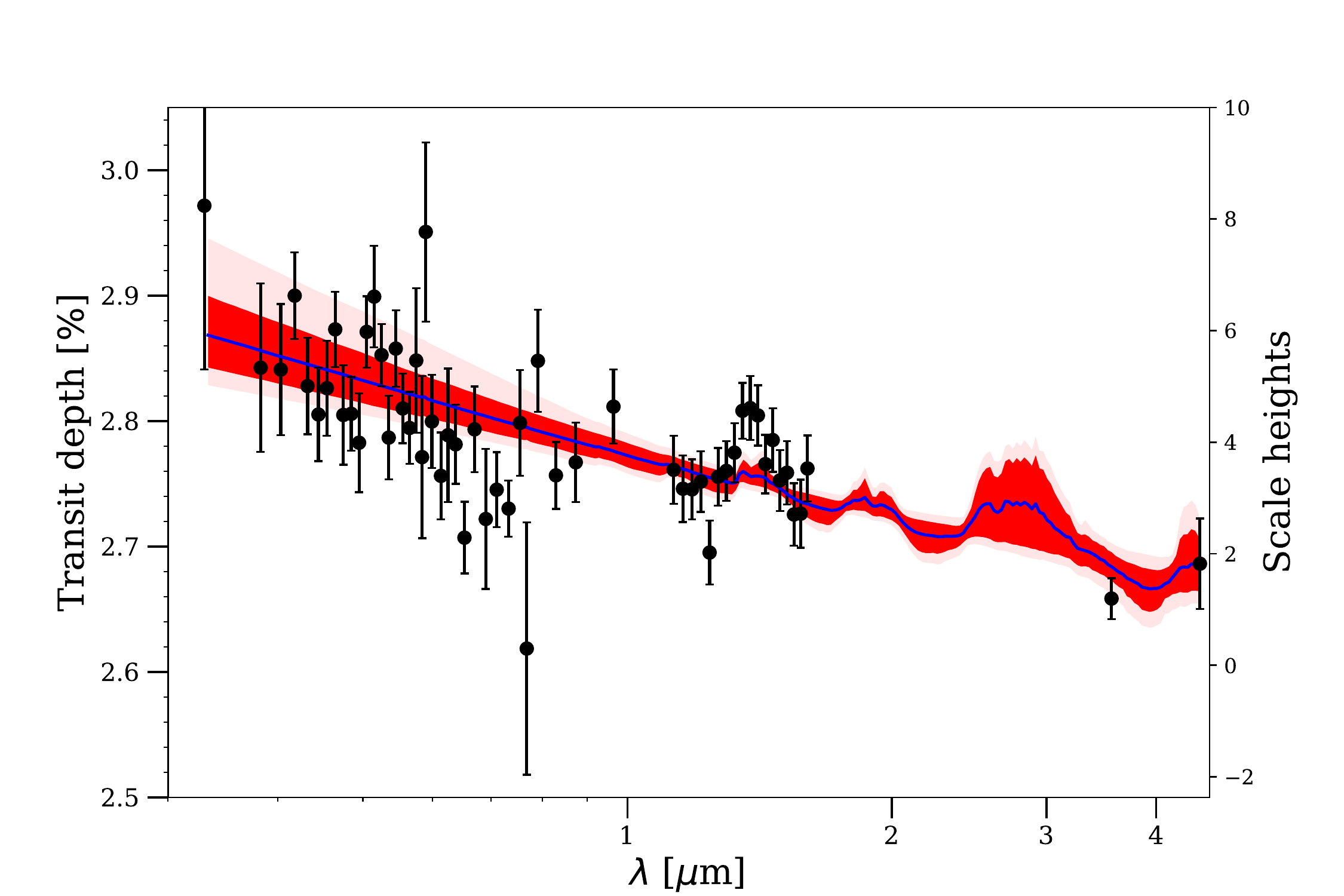}
\caption{Best solution and 1 and $2\sigma$ credible intervals for the \texttt{CHIMERA}-chemical equilibrium retrieval with a facula acting on the STIS data. To fit the visible spectrum, the needed increased scattering slopes mutes the WFC3-IR water feature.}
\label{stisretrieval}
\end{figure*}


\bsp	
\label{lastpage}
\end{document}